\documentclass[aps,pre,onecolumn,groupedaddress,longbibliography,nofootinbib,showkeys,showpacs,amssymb]{revtex4-1}

\usepackage{latexsym}
\usepackage{amsfonts}
\usepackage{graphicx}
\usepackage{mathptmx}      
\usepackage{bm}
\usepackage{enumerate}
\usepackage{subfigure}
\usepackage{color}
\usepackage{ulem}

\usepackage{amsmath,amssymb}
\usepackage{float}

{%
   \begin{list}{$\bullet$\ \ }%
   {%
      \setlength{\itemindent}{0pt}
      \setlength{\leftmargin}{0.5zw}%
      \setlength{\rightmargin}{0zw}%
      \setlength{\labelsep}{0.5zw}%
      \setlength{\labelwidth}{0zw}%
      \setlength{\itemsep}{0em}%
      \setlength{\parsep}{0em}%
      \setlength{\listparindent}{0zw}%
   }
}{%
   \end{list}%
}

\newcommand{\ignore}[1]{}
\newcommand{\beq}{\begin{equation}}
\newcommand{\eeq}{\end{equation}}

\newcommand{\ltsim}{\protect\raisebox{-0.5ex}{$\:\stackrel{\textstyle <}{\sim}\:$}}

\begin{document}

\title
{Computation of ESR spectra from the time evolution of the magnetization: comparison of autocorrelation and Wiener-Khinchin-relation based methods}

\author{Hiroki Ikeuchi$^{1,2}$}
\email[Corresponding author. Email address: ]{ikeuchi@spin.phys.s.u-tokyo.ac.jp}
\author{Hans De Raedt$^{3}$}
\author{Sylvain Bertaina$^{4}$}
\author{Seiji Miyashita$^{1,2}$}

\affiliation{$^{1}
${\it Department of Physics, Graduate School of Science,} The University of Tokyo, 7-3-1 Bunkyo-Ku, Tokyo, 113-0033, Japan \\
$^{2}${\it CREST, JST, 4-1-8 Honcho Kawaguchi, Saitama, 332-0012, Japan}\\
$^{3}${\it Department of Applied Physics, Zernike Institute for Advanced Materials,
University of Groningen, Nijenborgh 4, NL-9747AG Groningen, The Netherlands}\\
$^{4}${\it Aix-Marseille Universit\'e, CNRS, IM2NP UMR7334, F-13397 Marseille Cedex 20, France}
}

\date{\today}

\begin{abstract}
The calculation of finite temperature ESR spectra for concrete specified crystal configurations is a very important issue in the study of quantum spin systems.
Although direct evaluation of the Kubo formula by means of numerical diagonalization yields exact results,
memory and CPU-time restrictions limit the applicability of this approach to small system sizes.
Methods based on the time evolution of a single pure quantum state can be used to study larger systems.
One such method exploits the property that
the expectation value of the autocorrelation function obtained for a few samples of so-called 
thermal
typical states
yields a good estimate of the thermal equilibrium value.
In this paper, we propose a new method based on a Wiener-Khinchin-like theorem for quantum system.
By comparison with exact diagonalization results, it is shown that both methods yield correct results.
As the Wiener-Khinchin-based method involves sampling over 
thermal typical 
states,
we study the statistical properties of the sampling distribution.
Effects due to finite observation time are investigated and found to be different for the two methods
but it is also found that for both methods, the effects vanish as the system size increases.
We present ESR spectra of the one-dimensional XXZ Heisenberg chain of up to 26 spins
show that double peak structure due to the anisotropy is a robust feature of these spectra.
\end{abstract}


\maketitle


\section{Introduction}

Quantum spin systems have attracted interests for decades because they exhibit various nontrivial behavior due to quantum mechanical effects.
In particular, in low dimensions, quantum fluctuations due to non-commutativity of the spin operators
and/or competition among the interaction (frustration) play
an important
 role and various novel concepts,
such as the valence-bond solid, resonating-valence bonds, and magnon Bose-Einstein condensation, etc. have been developed.
One important topic is the effect of nonmagnetic defects.
In a spin $S=1/2$ antiferromagnetic Heisenberg chain,
quantum fluctuations prevent the spins from being ordered, even at $T=0$K,
and the ground state is nonmagnetic.
However, nonmagnetic defects break the translational symmetry and polarize the surrounding spins~\cite{R0,R1,R2}.
Then the one-dimensional system is described by an open-ended spin chain.
An example of such a system is the Pd doped chain Sr$_2$CuO$_3$\cite{R0exp,R3,R4}.

ESR is one of the major tools to study the effects of defects in spin systems.
Modeling the ESR spectra of intrinsic defects in spin chains is an important problem
for which data for finite but rather long chains are necessary.
In particular, the parameter dependence of concrete ESR spectrum for a specified system is of great interest
and the temperature dependence of the ESR spectrum provides a lot of information about the spin ordering.
To study these aspects theoretically, the explicit form of interactions and spatial configuration of magnetic ions in the lattice
play an important role and it is necessary to study microscopic models, that is
we should calculate the ESR spectrum for specific quantum spin Hamiltonian.
The most direct way is to calculate the Kubo formula~\cite{kubo_tomita,kubo}
by making use of the eigenvalues and their eigenvectors obtained by diagonalization of the Hamiltonian.
The
 first attempt has been made to study the Nagata-Tazuke phenomena~\cite{Nagata-Tazuke}
in a one-dimensional Heisenberg chain of 8 spins with dipole-dipole interactions~\cite{miyashita}.
In this work, the dependence of the spectrum on the angle between the static field and lattice direction was reproduced.
Moreover, antiferromagnetic resonance~\cite{ogasahara1} and also the appearance of a resonance forbidden by the
Dzyaloshinskii-Moriya interaction\cite{ogasahara2} were studied.
The structure of the ESR spectrum of a spin ring at high temperatures was investigated up to 16 spins~\cite{cepas}.
Complementary, Oshikawa and Affleck have developed a field theoretical approach by making use of the exact autocorrelation of Heisenberg chain,
and have successfully analyzed low temperature properties for long (infinite) chains~\cite{OA}.

Obviously, the application of this method of exact diagonalization (ED) is limited to small systems
for which we can obtain all the eigenvalues and their eigenvectors.
For a system of $N$ $S=1/2$ spins, we need the memory of order $2^{2N}$.
By making use of the symmetries of the system we may reduce the dimensions of the block diagonalized Hamiltonian.
Even with these efforts, in practice, the system size is still limited to about 20 spins.

The restriction imposed by the exact diagonalization approach can be alleviated by computing
the autocorrelation function (AC) from the time evolution of a pure state~\cite{vries,iitaka,machida}.The ESR spectrum is obtained by Fourier
 transformation
 of the autocorrelation function of the transverse magnetization.
In the AC method,
a
 pure state evolves in time according to the time-evolution operator $\mathrm{e}^{-i{\cal H}t}$
the action of which can be computed efficiently by means the Chebyshev polynomial~\cite{TALE84,LEFO91,IITA97,DOBR03}
or the Suzuki-Trotter-product-formula method~\cite{vries,hams}.
As the memory required to store one pure state or all eigenstates is of the order $2^N$ and $2^{2N}$, respectively,
the time evolution method allows us to study systems that are twice the size of those that can be studied by exact diagonalization.
For compute thermal averages we need, in principle, to take the average over the initial states.
However, it is known that the expectation value of a quantity $A$ for a single random state $|\Phi\rangle$
yields an estimate of the trace of $A$, that is $\langle \Phi|A|\Phi\rangle\simeq {\rm Tr}~A$,
which becomes more accurate as the size of Hilbert space increases~\cite{hams}.
Because of this fact, there is no need to compute traces of matrices to obtain the thermal equilibrium average~\cite{hams,iitaka,machida,raed00,shimizu}.
This approach has been used to study the temperature dependence of the total amplitude of ESR spectrum
for the single molecular magnet V$_{15}$, consisting of 15 $S=1/2$ spins~\cite{machida}.
Although this method to compute the ESR spectrum has considerable potential, 
properties of its applications to large systems
have not yet been scrutinized in sufficient detail.
An important issue, which we address in this paper, is that because
the time-evolution method necessarily yields data for a finite time interval only, it is
important to study how the ESR spectrum, that is the Fourier transform of this data, depends on the size of the time interval.

In the AC method, the spectrum is obtained from the autocorrelation function $\langle M^xM^x(t)\rangle_{\rm eq}$.
In ESR experiments, one measures the time evolution of the magnetization $M^x(t)$.
The relation between the AC and the spectral density $M^x(t)$ is given by the Wiener-Khinchin (WK) theorem.
This theorem relates the spectral density of the dynamics of a quantity and
the Fourier transform of the autocorrelation function of the same quantity.
In the present paper,  we propose a method to directly compute the ESR spectrum from the time evolution of $M^x(t)$ by exploiting the idea of WK theorem.
In quantum systems, however, the definition
of
 the magnetization dynamics is somewhat tricky
and we therefore develop a quantum version of Wiener-Khinchin relation,
i.e., an explicit relation between Fourier transform of the autocorrelation function and the spectrum density in the quantum case.
We call the approach based of this idea the Wiener-Khinchin (WK) method.
Because the thermal average of the magnetization in the transverse direction
$\langle M^x(t)\rangle_{\mathrm{eq}}$ 
is zero,
and the ESR signal is proportional to the average of square of the Fourier transform of
$\langle \Phi|M^x(t)|\Phi\rangle$ over many realizations of the random state
$|\Phi\rangle$
,the reasoning about the convergence of this average as a function of the dimension of the Hilbert space does not hold.
Therefore, we present a detailed analysis of the distribution of the sampled data,
and show that the distribution converges with a finite variance, independent of the size of Hilbert space.
Because of the latter property, also for a large system, it is necessary to perform ensemble averaging of the data
which renders the computational efficiency of the WK method less
than
the one of the AC method.
However, the WK method gives additional information about effects of finiteness of time domain.
In particular, the Gibbs oscillations in the WK method are positive only whereas they can be negative in the AC method.
We find that as the size of system increases, for both methods, the effects of finite time interval disappear
and the ESR spectra are very similar.
For large systems, the AC method is the most efficient one because a single sample of a random state yields accurate result.
We use both methods for applications to large systems which have not yet been studied in previous works.
In particular, we compute spectra for one-dimensional XXZ model up to 26 spins and show that
the double peak which was found in earlier work~\cite{cepas} is almost independent of the system size,
indicating that the double-peak structure will be present in the thermodynamic limit.

The outline of this paper is as follows.
A brief overview on the methods previously used is given in Sec~\ref{sec_method}.
In Sec.~\ref{sec_WK}, we introduce the new method motivated
by the Wiener-Khinchin theorem, and
study statistical properties of the method in detail in Sec.~\ref{sec_variance}.
In Sec.~\ref{sec_dist}, we show the distribution of sampled data of the spectrum
 obtained by the AC and WK methods.
The total amplitude of the spectrum, which is fundamental information in ESR studies, is discussed in Sec.~\ref{amplitude}.
Section~\ref{sec_APPLICATION} presents applications of the methods to large systems.
Summary and discussion of related problems are given in Sec.~\ref{sec_summary}.
Appendix A discusses the effect of the finite time interval on the spectrum for the both methods and
Appendix B gives a detailed analysis of the statistical properties of the thermal typical state for the both methods.

\section{Recapitulation of existing methods}\label{sec_method}

\subsection{Kubo formula and ESR spectrum} 
The ESR spectrum is given by the Kubo formula\cite{kubo_tomita,kubo}.
The imaginary part of the dynamical susceptibility $\chi''(\omega)$ reads\footnote{%
Throughout the paper we denote the temperature by $\beta^{-1}$ in order to avoid confusion with the end of the time interval $T$.}
\beq
	\chi''(\omega)=
	\frac{1}{2}(1-\mathrm{e}^{-\beta\omega})\int_{-\infty}^{\infty}\langle M^{x}(0)M^{x}(t) \rangle_{\mathrm{eq}}\mathrm{e}^{-\mathrm{i}\omega t}\mathrm{d}t,
\label{eq:chi}
\eeq
and the ESR absorption spectrum is given by
\beq
I^{x}(\omega)=\frac{\omega\lambda_{0}^{2}}{2}\chi''(\omega),
\label{Ixomega}
\eeq
where $\lambda_{0}$ is the amplitude of the external field, and we adopt
\begin{eqnarray}
      	M^{x}(t)=\mathrm{e}^{\mathrm{i}\mathcal{H}t}M^{x}\mathrm{e}^{-\mathrm{i}\mathcal{H}t},\quad M^{x}=\sum_{i=1}^{N}S_{i}^{x},\\
	\langle \cdot \rangle_{\mathrm{eq}}=\mathrm{Tr}[\hspace{1mm}\cdot\hspace{1mm}\mathrm{e}^{-\beta\mathcal{H}}]/\mathrm{Tr}[\mathrm{e}^{-\beta\mathcal{H}}].
\end{eqnarray}

For the numerical calculation the ESR absorption spectrum, several methods have been developed.
There are essentially two types of methods:
 (1) Exact diagonalization method~\cite{miyashita,cepas}  and (2) Time-evolution of the autocorrelation function method~\cite{iitaka}.

\subsection{Exact diagonalization (ED) method}

The most direct calculation of the Kubo formula uses the set of eigenvalues and eigenvectors
$\{{{E_{n},|n\rangle}\}_{n=1}^{D}}$ obtained by solving the eigenvalue problem
\beq
\mathcal{H}|n\rangle=E_{n}|n\rangle
\label{eigenstateH}
\eeq
for the hamiltonian $\mathcal{H}$, $D$ denoting the dimension of the Hilbert space.
In practice, we solve Eq.~(\ref{eigenstateH}) by numerical diagonalization.
The autocorrelation function $\langle M^{x}(0)M^{x}(t)\rangle_{\mathrm{eq}}$ is expressed as
\begin{eqnarray}
	 \langle M^{x}(0)M^{x}(t)\rangle_{\mathrm{eq}}
	&=&\sum_{m,n}|\langle m|M^{x}|n\rangle|^{2}\mathrm{e}^{\mathrm{i}(E_{m}-E_{n})t-\beta E_{n}}/Z,
       \label{AC0}
\end{eqnarray}
where $Z$ is the partition function
\beq
Z=\sum_{n}\mathrm{e}^{-\beta E_{n}}.
\eeq
The Fourier transform of Eq.~(\ref{AC0})  reads
\begin{eqnarray}
	\int_{-\infty}^{\infty}\langle M^{x}(0)M^{x}(t)\rangle_{\mathrm{eq}}\mathrm{e}^{-\mathrm{i}\omega t}\mathrm{d}t
	&=&\sum_{m,n}|\langle m|M^{x}|n\rangle|^{2}\mathrm{e}^{-\beta E_{n}}2\pi\delta\left(\omega-(E_{m}-E_{n})\right)/Z,
\end{eqnarray}
where we used the definition
\begin{eqnarray}
\delta\left(\omega-(E_{m}-E_{n})\right)={1\over 2\pi}
	\int_{-\infty}^{\infty}\mathrm{e}^{-\mathrm{i}\left(\omega-(E_{m}-E_{n})\right)t}\mathrm{d}t.
\end{eqnarray}
The imaginary part of the dynamical susceptibility reads
\begin{eqnarray}
	\chi''(\omega)
	&\equiv&\sum_{m,n}D_{m,n}\delta\left(\omega-\omega_{m,n}\right),
	\label{chidelta}
\end{eqnarray}
where
\begin{eqnarray}
	D_{m,n}\equiv\pi(\mathrm{e}^{-\beta E_{n}}-\mathrm{e}^{-\beta E_{m}})|\langle m|M^{x}|n\rangle|^{2}/Z,\quad \omega_{m,n}\equiv E_{m}-E_{n}.
\end{eqnarray}
It is sufficient to consider $\omega_{m,n}>0$ only since we are interested in the absorption, not the emission.
Note that $\chi''(\omega)>0$ for $\omega>0$.

In the exact diagonalization approach, the spectrum consists of a finite sum of delta functions.
Therefore, to draw the spectrum, we have to represent each delta function by a regular function of certain width.
For example, we may replace the delta function by a Gaussian or
simply
 use a histogram representation.
The results of the exact diagonalization are exact, usually close to machine precision,
but we need to know all the eigenvalues and corresponding eigenstates.
Therefore, we need memory of the order of 
$D^2$
for a matrix of the size $D=2^N$ for systems of $N$ $S=1/2$ spins.
If the system has symmetry, we may reduce the size.
For example if the system conserves $M^{z}$, then $D$ is reduced to $_NC_{N/2+M^z}$.
Moreover for ESR spectra, only the uniform mode is relevant and therefore only the fully symmetrized states are necessary,
which also allows a reduction of $D$.
However, the memory limitation prevents us from studying more that $N=20$ $S=1/2$ spins.
As an illustration, we use the exact diagonalization approach to study
the one-dimensional spin-1/2 XXZ model in a static magnetic field $H$ and compute
the response of the magnetization along the $x$ axis to the AC-field $\lambda(t)$.
The hamiltonian of the system is given by
\begin{eqnarray}
	\mathcal{H_{\mathrm{tot}}}&=&\sum_{i=1}^{N}(J_xS_{i}^{x}S_{i+1}^{x}+J_yS_{i}^yS_{i+1}^y+J_zS_{i}^zS_{i+1}^z)
	-g\mu_{\rm B}H\sum_{i=1}^{N}S_{i}^{z}
	+\lambda_{0}\mathrm{cos}\omega t\sum_{i=1}^{N}S_{i}^{x},
	\label{eq:XXZ}
\end{eqnarray}
where we impose the periodic boundary condition $S_{N+1}^{\alpha}=S_{1}^{\alpha},\,\alpha=x,y,z$ unless otherwise mentioned.
Hereafter we adopt Kelvin as the unit of energy and we set $g\mu_{\rm B}=1$.

\begin{figure}[H]
	\begin{center}
	\includegraphics[width=100mm]{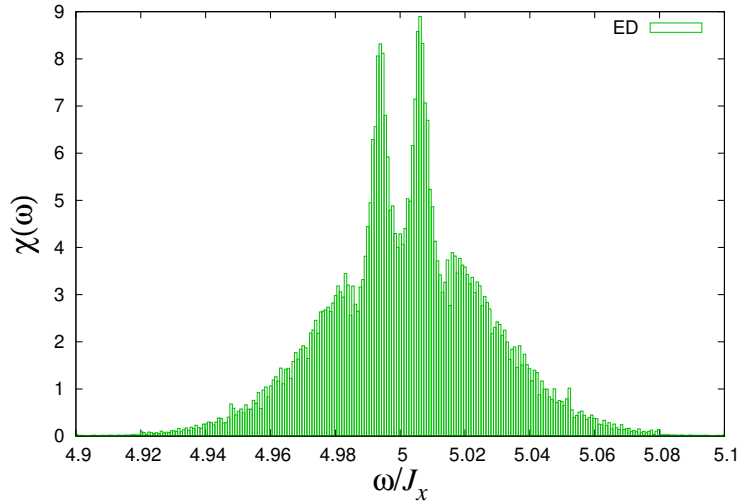}
	\end{center}
	\caption{Spectrum for a ring of $N=16$ spins. The histogram is obtained by the exact diagonalization method.}
	\label{N16ED}
\end{figure}

In Fig.~\ref{N16ED}, we show an example of spectrum obtained by exact diagonalization method for an antiferromagnetic Heisenberg chain with $N=16$ with the static field
$H=5\mathrm{K}$
at the temperature $\beta^{-1}=100\mathrm{K}$. The notation and the parameters of the model are set to be the same as those in the model studied in the previous study\cite{cepas}  $(J_x=J_y=1\mathrm{K},J_z=0.92\mathrm{K})$.

The spectrum consisting of an ensemble of the delta peaks (Eq.~(\ref{chidelta})) is plotted as
a histogram with small bins $\Delta\omega$.
In the exact diagonalization (ED) method, we are free to choose any value of $\Delta\omega$.
Anticipating for the comparison with the two other methods,
we take $\Delta\omega=2\pi/T$ where $T=\mathrm{d}t\times n_t$, $\mathrm{d}t=0.5$ and $n_t=16384$ is the number of data points.

\subsection{Autocorrelation function method}
\label{sec:TDACFN}

According to the Kubo formula (Eq.(\ref{eq:chi})), the spectrum is given by Fourier transform of the autocorrelation function.
Because of the symmetry
\beq
\left(\langle M^{x}M^{x}(t)\rangle_{\mathrm{eq}}\right)^{*}=\langle M^{x}M^{x}(-t)\rangle_{\mathrm{eq}},
\label{MMtstar}
\eeq
where $*$ denotes complex 
conjugate,
 we have
\begin{eqnarray}
\label{AC_integration}
\int_{-\infty}^{\infty}\langle M^{x}M^{x}(t)\rangle_{\mathrm{eq}}\mathrm{e}^{-\mathrm{i}\omega t}\mathrm{d}t
=2\mathrm{Re}\left[\int_{0}^{\infty}\langle M^{x}M^{x}(t)\rangle_{\mathrm{eq}}\mathrm{e}^{-\mathrm{i}\omega t}\mathrm{d}t\right],
\end{eqnarray}
where $\mathrm{Re[\cdot]}$ denotes the real part.
Therefore, to compute Eq.~(\ref{AC_integration}), it suffices to have AC data for the interval $[0,T]$.

We obtain the spectrum by discrete Fourier
 transform
 (DFT) of the AC data
\beq
f(t_m)\equiv\langle M^{x}M^{x}(t_m)\rangle_{\mathrm{eq}},\quad t_m={mT\over n_t},\quad m=0,1,2,\cdots, n_t-1,
\eeq
where $n_t$ is the number of data items.
The DFT of Eq.~(\ref{AC_integration}) is given by
\begin{eqnarray}
	\frac{T}{n_t}\sum_{m=-n_t}^{n_t-1}f\left(\frac{mT}{n_t}\right)\mathrm{e}^{-\mathrm{i}\omega_{k}mT/{n_t}},\quad
	\omega_{k}\equiv\frac{\pi k}{T},\quad k=-n_t, -n_t+1,\cdots, 0,1,2,\cdots,n_t-1.
       \label{DFT}
\end{eqnarray}
Note that the absorption and emission spectrum correspond to $\omega_k >0$ and $\omega_k <0$, respectively.
The maximum angular frequency ($2\pi$ times Nyquist frequency) 
that can be represented by the DFT Eq.~(\ref{DFT}) is given by $\pi n_t/{T}$.
For a magnetic field $H=5\mathrm{K}$, the main contribution to the absorption spectrum peak comes from a narrow peak around
$\omega\approx 5\mathrm{K}$. In our numerical work we take $\mathrm{d}t\equiv T/n_t=0.5$.
Therefore the largest value of omega ($2\pi$ times the Nyquist frequency) is
$(2\pi/T)\times (n_t/2)=\pi/\mathrm{d}t=2\pi\approx 6.28 >5$, which is large enough to
cover the full spectral range.

\begin{figure}[H]
	\begin{center}
		{\includegraphics[width=80mm]{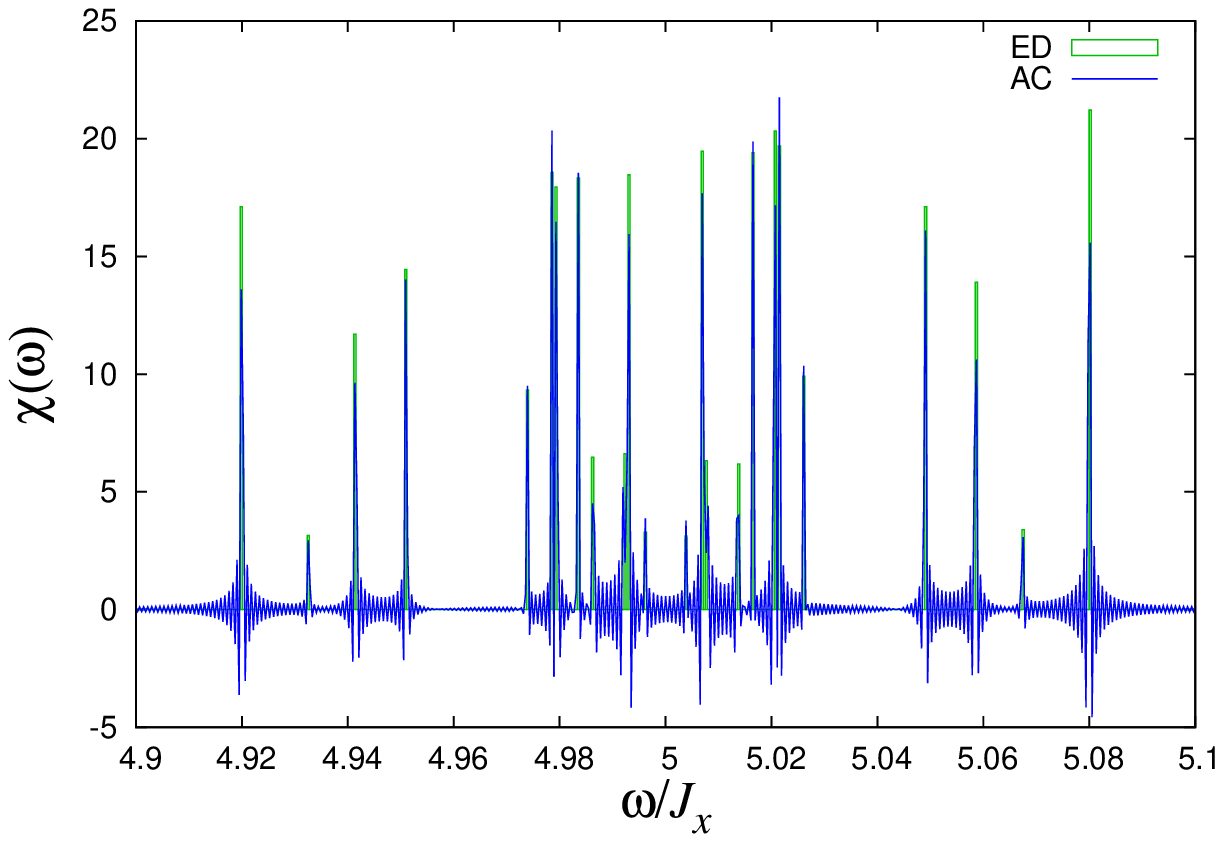}}
		{\includegraphics[width=80mm]{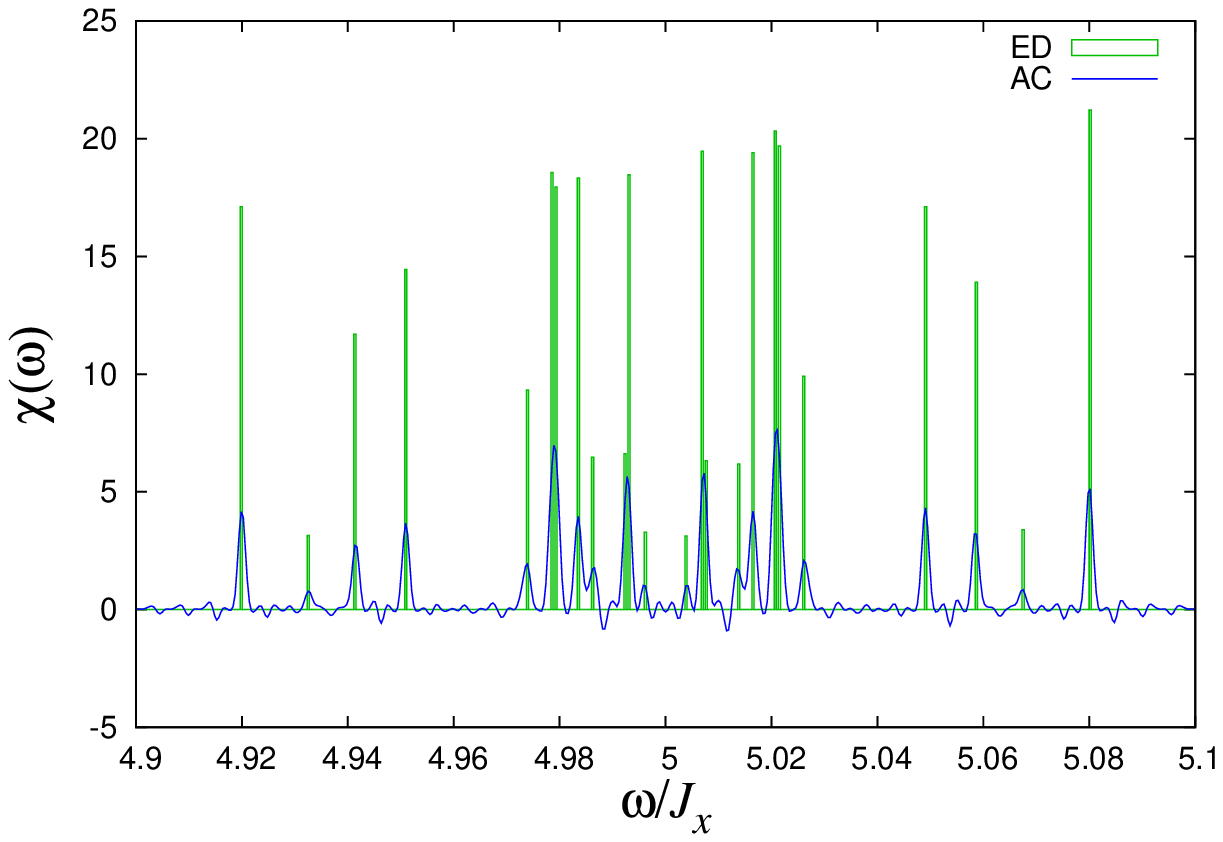}}
	\end{center}
%
\caption{The spectrum obtained by exact diagonalization method and by the time-evolution method:
$N=6$, temperature $\beta^{-1}=100\mathrm{K}, H=5\mathrm{K}$,
sampling time $\mathrm{d}t=0.5$, the number of data items $n_t=16384$,
and the $\omega$-mesh $\pi/T=\pi/(n_t\times\mathrm{d}t)$ (blue line).
The exact diagonalization result (green line), uses an $\omega$-mesh $\Delta\omega=2\pi/16384$,
which corresponds to the frequency $\pi/T$ used in time-evolution method.
Left: the spectrum obtained by exact diagonalization (green line)
and by time-evolution method without window functions (blue line).
Strong 
Gibbs oscillation is found.
Right: Same as left except that a Gaussian window function
with standard deviation ${\alpha}/{T}=7/8192=0.00085$ ($\alpha = 7$) is used.
Clearly, the 
Gibbs oscillation is suppressed.
}
\label{EDTE}
\end{figure}

In Fig.~\ref{EDTE}(left), we show the spectrum for a system of $N=6$ spins at the temperature $\beta^{-1}=100$K, obtained by using a time series
of $n_t=16384$ items.
Clearly, the spectrum is suffering from fine oscillations with negative values
values,
 which is called the Gibbs oscillation\cite{harris}.
This oscillation is due to the fact that the range of the time integral is finite and that there are eigenvalues that do not
not exactly match one of the $\omega_{k}$.
This artifact can be suppressed by employing a suitable window function\cite{harris}, for instance a Gaussian window (see Appendix A).
As shown in Fig.~\ref{EDTE} (Right), the Gibbs oscillation is reduced and the result is consistent
with the one derived with the exact diagonalization method with the resolution of $O(\alpha)$
($\alpha$ is an artificial parameter which determines the resolution of the spectrum. See Appendix A for the precise definition).

\subsubsection{Thermal typical state method} 

In theory, the calculation of the thermal equilibrium expectation value of $M^xM^x(t)$ involves a trace operation.
This requires of the order $D^2=2^{2N}$ operations, which becomes prohibitively large if $N$ is of the order of 20 or larger.
Fortunately, for larger systems, we can obtain accurate estimates of thermal equilibrium averages by making use of
the so-called thermal typical state 
$|\Phi_{\beta}\rangle=\mathrm{e}^{-\beta \mathcal{H}/2}|\Phi\rangle$
~\cite{hams,iitaka,machida,shimizu}, also called
``a Boltzmann-weighted random vector''~\cite{iitaka,machida} or ``a canonical thermal pure quantum state''\cite{shimizu},
where $|\Phi\rangle$ denotes a random state on the $D$-dimensional 
hypershere.
In essence, we have~\cite{hams}
\beq
\langle X\rangle_{\rm eq}=\frac{\mathbf{Tr\;} Xe^{-\beta {\cal H}} }{\mathbf{Tr\;} e^{-\beta {\cal H}} }
\simeq \langle \Phi_{\beta}|X|\Phi_{\beta}\rangle.
\eeq
We briefly explain the ideas behind this method.
Let $\{|n\rangle\}_{n=1}^{D}$ be an arbitrary set of complete orthonormal states of the Hilbert space of the system.
Using the complex-valued random variables $\{\xi_{n}\}_{n=1}^{D}$, we introduce a random vector:
\begin{eqnarray}
	|\Phi\rangle=\sum_{n=1}^{D}\xi_{n}|n\rangle.
\end{eqnarray}
Note that this construction is independent of the choice of the basis set $\{|n\rangle\}_{n=1}^{D}$.

Unlike in 
[A. Hams and H. De Raedt, PRE (2000)]
, to simplify the mathematics,
we will work with independent complex-valued Gaussian random variable for each index $n$.
They have identical Gaussian distributions with mean zero and variance $2\sigma^2$ for all real
and imaginary parts of the variables.
By $\mathrm{E}[\cdot]$,
we denote the expectation with respect to the multivariate Gaussian probability distribution:
\begin{eqnarray}
P(\xi_{1},\ldots,\xi_{D})\prod_{a=1}^D\mathrm{d}(\mathrm{Re\;}\xi_a)\mathrm{d}(\mathrm{Im\;}\xi_a)&=&\prod_{a=1}^D \left[ \frac{1}{2\pi\sigma^2}\mathrm{e}^{-|\xi_{a}|^2/2\sigma^2}\right]
\mathrm{d}(\mathrm{Re\;}\xi_a)\mathrm{d}(\mathrm{Im\;}\xi_a).
\label{Gprob}
\end{eqnarray}
The following relations hold:
\begin{eqnarray}
\mathrm{E}[\xi_{a} ]&=&\mathrm{E}[\xi_{p}^{\phantom{^\ast}} ]=\mathrm{E}[\xi_{a}\xi_{b} ]=0,
\nonumber \\
\mathrm{E}[\xi_{a}^\ast \xi_{p}^{\phantom{^\ast}}]&=&2\sigma^2\delta_{a,p},
\nonumber \\
\mathrm{E}[\xi_{a}^\ast \xi_{b}^\ast \xi_{p}^{\phantom{^\ast}}\xi_{q}^{\phantom{^\ast}}]&=&
\mathrm{E}[\xi_{a}^\ast \xi_{p}^{\phantom{^\ast}}] \mathrm{E}[\xi_{b}^\ast \xi_{q}^{\phantom{^\ast}}]
+
\mathrm{E}[\xi_{a}^\ast \xi_{q}^{\phantom{^\ast}}] \mathrm{E}[\xi_{b}^\ast \xi_{p}^{\phantom{^\ast}}]
=4\sigma^4\left(\delta_{a,p}\delta_{b,q}+ \delta_{a,q}\delta_{b,p}\right),
\label{RS30text}
\end{eqnarray}
which are used in the following calculations.

For any matrix $X$ we have
\begin{eqnarray}
\mathrm{E}[\langle\Phi|X|\Phi\rangle]
&=&
\sum_{a,p=1}^D \mathrm{E}[ \xi_a^\ast \xi_p^{\phantom{\ast}} ] \langle a|X|p\rangle
=\sum_{a}^D \mathrm{E}[ \xi_a^\ast \xi_a^{\phantom{\ast}} ] \langle a|X|a\rangle
=2\sigma^2 \mathbf{Tr\;} X,
\label{RS31text}
\end{eqnarray}
and because $\langle\Phi|X^{\dagger}|\Phi\rangle=\langle\Phi|X|\Phi\rangle^\ast$, the corresponding variance is given by
\begin{eqnarray}
\mathrm{Var}\left(\langle\Phi|X|\Phi\rangle\right)
&=&\mathrm{E}[|\langle\Phi|X|\Phi\rangle|^2]-\left|\mathrm{E}[\langle\Phi|X|\Phi\rangle]\right|^2
\nonumber \\
&=&\sum_{a,p,b,q=1}^D \mathrm{E}[ \xi_a^\ast \xi_p^{\phantom{\ast}} \xi_b^{\phantom{\ast}} \xi_q^\ast ]
\langle a|X|p\rangle
\langle b|X|q\rangle^\ast - 4\sigma^4
\left|\mathbf{Tr\;} X\right|^2
\nonumber \\
&=&4\sigma^4 \sum_{a,b=1}^D
\bigg(\langle a|X|a\rangle
\langle b|X|b\rangle^\ast
+
\langle a|X|b\rangle
\langle a|X|b\rangle^\ast
\bigg)
- 4\sigma^4 \left|\mathbf{Tr\;} X\right|^2
\nonumber \\
&=&4\sigma^4
\mathbf{Tr\;} XX^\dagger
.
\label{RS32text}
\end{eqnarray}
The relative standard deviation is defined by
\begin{eqnarray}
\mathrm{RSD}^2(\langle\Phi|X|\Phi\rangle)&=&\frac{\mathrm{Var}(\langle\Phi|X|\Phi\rangle)}{
|\mathrm{E}[\langle\Phi|X|\Phi\rangle]|^2}
= \frac{\mathbf{Tr\;} XX^\dagger}{\left|\mathbf{Tr\;} X\right|^2}
= \frac{\mathbf{Tr\;} XX^\dagger}{(\mathbf{Tr\;} X) (\mathbf{Tr\;} X^\dagger) }.
\label{RS34text}
\end{eqnarray}

Next, we introduce the thermal typical state $|\Phi_{\beta}\rangle$ for the system at an inverse temperature $\beta$:
\begin{eqnarray}
	|\Phi_{\beta}\rangle=\mathrm{e}^{-\beta\mathcal{H}/2}|\Phi\rangle,
\end{eqnarray}
where $\langle\Phi_{\beta}|\Phi_{\beta}\rangle=\langle\Phi|\mathrm{e}^{-\beta\mathcal{H}}|\Phi\rangle$ is an approximation to the partition function $Z_{\beta}$.
As shown in Appendix B, the RSD of $\langle\Phi|\mathrm{e}^{-\beta\mathcal{H}}|\Phi\rangle$ vanishes with $D$ as
\begin{eqnarray}
\mathrm{RSD}(\langle\Phi_{\beta}|\Phi_{\beta}\rangle)=
\mathrm{RSD}(\langle\Phi|\mathrm{e}^{-\beta\mathcal{H}}|\Phi\rangle)\ltsim\mathcal{O}(D^{-1/2}).
\end{eqnarray}
Thus, hereafter we only estimate the RSD of the numerator for the thermal average.
In our calculations, we take averages of the dominator and the numerator with respect to the same samples $\{|\Phi_{\beta}\rangle\}$:
\begin{eqnarray}
\langle X\rangle_{\rm eq}\simeq{\frac{1}{N}\sum_{|\Phi_{\beta}\rangle}\langle\Phi_{\beta}|X|\Phi_{\beta}\rangle\over \frac{1}{N}\sum_{|\Phi_{\beta}\rangle}\langle\Phi_{\beta}|\Phi_{\beta}\rangle},
\end{eqnarray}
where $N$ is the number of samples.
But if we take the average of the ratios of samples, we obtain the same result in the limit $D\rightarrow\infty$ (see Appendix B):
\begin{eqnarray}
\langle X\rangle_{\rm eq}\simeq\frac{1}{N}\sum_{|\Phi_{\beta}\rangle}{\langle\Phi_{\beta}|X|\Phi_{\beta}\rangle\over \langle\Phi_{\beta}|\Phi_{\beta}\rangle}.
\end{eqnarray}
The autocorrelation function $\langle M^{x}M^{x}(t)\rangle_{\mathrm{eq}}$ is given by
\begin{eqnarray}
	 \langle M^{x}M^{x}(t)\rangle_{\mathrm{eq}}=\frac{\mathrm{E}[\langle\Phi_{\beta}|M^{x}\mathrm{e}^{\mathrm{i}\mathcal{H}t}M^{x}\mathrm{e}^{-\mathrm{i}\mathcal{H}t}|\Phi_{\beta}\rangle]}{\mathrm{E}[\langle\Phi_{\beta}|\Phi_{\beta}\rangle]}.
	 \label{TREthermal}
\end{eqnarray}
The RSD of the autocorrelation function also vanishes with increasing $D$ (see Appendix B).
Thus, for a sufficiently large system, a few samples suffice to estimate the autocorrelation function.
Here it should be noted that strictly speaking Eq.~(\ref{MMtstar}) does not hold for the quantity
$\langle\Phi_{\beta}|M^{x}\mathrm{e}^{\mathrm{i}\mathcal{H}t}M^{x}\mathrm{e}^{-\mathrm{i}\mathcal{H}t}|\Phi_{\beta}\rangle$,
but for sufficiently large $D$, Eq.~(\ref{TREthermal}) justifies using Eq.~(\ref{MMtstar}).

For the autocorrelation function $\langle M^{x}M^{x}(t)\rangle_{\mathrm{eq}}$, we need to construct the states
\beq
|A\rangle\equiv \mathrm{e}^{-\mathrm{i}{\cal H}t}|\Phi_{\beta}\rangle,\quad {\rm and}\quad
|B\rangle\equiv \mathrm{e}^{-\mathrm{i}{\cal H}t}M^{x}|\Phi_{\beta}\rangle,
\eeq
and then obtain the expectation value of the autocorrelation as
\beq
\langle\Phi_{\beta}|M^{x}\mathrm{e}^{\mathrm{i}\mathcal{H}t}M^{x}\mathrm{e}^{-\mathrm{i}\mathcal{H}t}|\Phi_{\beta}\rangle=\langle B|M^{x}|A\rangle.
\eeq

To perform the operations $\mathrm{e}^{-\beta\mathcal{H}/2}$ and  
$\mathrm{e}^{-\mathrm{i}{\cal H}t}$
 on the states,
we may use the Chebyshev method or the Suzuki-Trotter decomposition method.
In the present study we used the Chebyshev method for  $\mathrm{e}^{-\beta\mathcal{H}/2}|\Phi\rangle$
and mostly used the product method for $\mathrm{e}^{-\mathrm{i}\mathcal{H}t}|\Phi_{\beta}\rangle$.

\section{Wiener-Khinchin method}\label{sec_WK} 

In this section, we propose a new method that makes use of the Wiener-Khinchin theorem.
In the autocorrelation method (see previous section) the spectrum is estimated from the autocorrelation function $\langle M^xM^x(t)\rangle_{\rm eq}$.
In experiments, the spectrum is obtained from the record of time evolution of magnetization $M^x(t)$.
The relation between the autocorrelation and the Fourier transform of time evolution of a quantity $X(t)$
is given by the Wiener-Khinchin relation, that is,
the relation between the spectrum of the fluctuation in time of a quantity $X(t)$
and the Fourier transform of the autocorrelation function  $\langle X(0)X(t)\rangle$.

We explore the possibility to obtain the spectrum from the dynamics of the quantity itself, i.e.
$\langle\Phi_{\beta}|X(t)|\Phi_{\beta}\rangle$ from an initial state $|\Phi_{\beta}\rangle$.
In this method, we use the time-evolution of the state as in the previous subsection
and therefore we can study large systems as well.

\subsection{Wiener-Khinchin theorem}

First, we briefly review the Wiener-Khinchin theorem.
For a time-dependent quantity $X(t)$, we define the autocorrelation function $R(t)$ and the spectral density $S(\omega)$ as
\beq
	R(t)\equiv\langle X(0)X(t)\rangle\equiv\lim_{T \to \infty}\frac{1}{T}\int_{0}^{T}X(\tau)X(t+\tau)\mathrm{d}\tau,
	\label{Rt}
\eeq
and
\beq
	S(\omega)\equiv\lim_{T \to \infty}\frac{|X^{T}(\omega)|^2}{T},\quad X^{T}(\omega)\equiv\int_{0}^{T}X(t)\mathrm{e}^{-\mathrm{i}\omega t}\mathrm{d}t,
	\label{Somega}
\eeq
respectively.
The Wiener-Khinchin theorem tells us that the Fourier transform of the autocorrelation function equals to spectral density:
\begin{eqnarray}
	G(\omega)\equiv\int_{-\infty}^{\infty}R(t)\mathrm{e}^{-\mathrm{i}\omega t}\mathrm{d}t=S(\omega).
	\label{Gomega}
\end{eqnarray}
Note that we assume that the process $X(t)$ is stationary\cite{vankampen}.

\subsection{Dynamics of the magnetization}

Next, we apply the idea of Wiener-Khinchin relation to the Kubo formula.
By making use of the relation (\ref{Gomega}) we obtain $G(\omega)$ from the spectral density $S(\omega)$.
Here it should be noted that $M^{x}(t)$ is an operator and the definition of $\langle M^{x}(t)\rangle$ is tricky.
Definitely, $\langle M^{x}(t)\rangle_{\mathrm{eq}}$ is time-independent, in fact zero in the present case, and therefore we cannot extract $S(\omega)$.

In the following, we propose a method to obtain $G(\omega)$ from $\langle M^{x}(t)\rangle$ in a quantum mechanical system.
This approach is motivated by the Wiener-Khinchin theorem but we do not use the relation (\ref{Gomega}) directly.
In time-domain methods, we use the notion of the thermal typical state (see above)
and we use this notion here once more to derive a formula to obtain $S(\omega)$
from time evolution of $\langle M^{x}(t)\rangle$~\cite{discuss}.

First we prepare 
a thermal typical
state
$|\Phi_{\beta}\rangle$ as an initial state.
The expectation value with this state gives the thermal average at the inverse temperature $\beta$.
For  $M^{x}(t)$,  $\langle M^{x}(t)\rangle_{\mathrm{eq}}=0$ and the expectation value
$\mathrm{E}[\langle\Phi_{\beta}| M^{x}(t)|\Phi_{\beta}\rangle]$ is zero,
but, in general, $\langle\Phi_{\beta}| M^{x}(t)|\Phi_{\beta}\rangle$ is not zero.
When we calculate the time evolution of the sampled state we can extract information for the spectral density
\beq
\hat{M}_{\beta}^{T}(\omega)\equiv \int_{0}^{T} \langle\Phi_{\beta}| M^{x}(t)|\Phi_{\beta}\rangle
\mathrm{e}^{-\mathrm{i}\omega t}\mathrm{d}t.
\label{mbetaTomega0}
\eeq
Although $E[\langle\Phi_{\beta}| \hat{M}^{x}(t)|\Phi_{\beta}\rangle]$ is zero,
the expectation value of the spectral density $\mathrm{E}[|\hat{M}_{\beta}^{T}(\omega)|^2]$
is not zero and, as we show below, we can obtain the ESR spectrum from the latter quantity.

Taking the basis to be the eigenstates of the system,
$\langle\Phi_{\beta}|M^{x}(t)|\Phi_{\beta}\rangle$ is expressed in tems of the random numbers $\{\xi_{n}\}$ as
\begin{eqnarray}
	\langle\Phi_{\beta}|M^{x}(t)|\Phi_{\beta}\rangle
	=\sum_{m,n}\xi_{m}^{*}\xi_{n}\mathrm{e}^{-\beta(E_{m}+E_{n})/2}\mathrm{e}^{\mathrm{i}(E_{m}-E_{n})t}
	\langle m|M^{x}|n\rangle,
	\label{mxbetat}
\end{eqnarray}
where
$|n\rangle$ and $E_n$ are the eigenvector and its eigenenergy of the system Hamiltonian ${\cal H}$ (Eq.~(\ref{eigenstateH})).
The Fourier transform ${\hat M}_{\beta}^{T}(\omega)$ (Eq.(\ref{mbetaTomega0})) is expressed as
\begin{eqnarray}
	{\hat M}_{\beta}^{T}(\omega)
=\sum_{m,n}\xi_{m}^{*}\xi_{n}\mathrm{e}^{-\beta(E_{m}+E_{n})/2}\mathrm{e}^{-\mathrm{i}(\omega-(E_m-E_n))T/2}2\pi\delta^{T}\left(\omega-(E_{m}-E_{n})\right)
	\langle m|M^{x}|n\rangle,
	\label{mxbetaomega}
\end{eqnarray}
where we introduced the function
\begin{eqnarray}
\delta^{T}(\omega)\equiv\frac{\mathrm{sin\frac{\omega T}{2}}}{\pi\omega}\xrightarrow{T \to \infty}\delta(\omega).
\end{eqnarray}
Note that we cannot extend the range of integration as in the AC method
(Eq.~(\ref{AC_integration})) because for a given $|\Phi_{\beta}\rangle$,
$\langle\Phi_{\beta}| M^{x}(t)|\Phi_{\beta}\rangle^*=\langle\Phi_{\beta}| M^{x}(-t)|\Phi_{\beta}\rangle$ does not hold in general.
For the same number of time steps, this implies that the frequency resolution  $\Delta\omega=2\pi/T$ is twice as larger as for the AC method.

As a next step, we calculate the average over the random initial states of the quantity
\begin{multline}
	|{\hat M}_{\beta}^{T}(\omega)|^2=\sum_{m,n}\sum_{m',n'}\xi_{m}^{*}\xi_{n}\xi_{m'}\xi_{n'}^{*}
	\mathrm{e}^{-\beta(E_{m}+E_{n})/2}\mathrm{e}^{-\beta(E_{m'}+E_{n'})/2}
	\times\mathrm{e}^{-\mathrm{i}(\omega-(E_m-E_n))T/2}\mathrm{e}^{+\mathrm{i}(\omega(E_{m'}-E_{n'}))T/2}\\
	\times4\pi^{2}\delta^{T}\left(\omega-(E_{m}-E_{n})\right)\delta^{T}\left(\omega-(E_{m'}-E_{n'})\right)
	\langle m|M^{x}|n\rangle\langle n'|M^{x}|m'\rangle.
	\label{Zsquare}
\end{multline}
Using Eq.~(\ref{RS30text}) and the fact that $\langle n|M^{x}|n\rangle$=0, we have
\begin{eqnarray}
	\label{exactWK}
	\mathrm{E}[|{\hat M}_{\beta}^{T}(\omega)|^2]
	&=&2\sigma^2T\mathrm{e}^{-\beta\omega}\sum_{m,n}\mathrm{e}^{-2\beta E_{n}}2\pi
	\delta^{T}\left(\omega-(E_{m}-E_{n})\right)|\langle m|M^{x}|n\rangle|^2,
\end{eqnarray}
where we used the relation
\begin{eqnarray}
	\left(\delta^{T}(\omega-(E_{m}-E_{n}))\right)^2\approx\frac{T}{2\pi}\delta^{T}
	\left(\omega-(E_{m}-E_{n})\right).
\end{eqnarray}
The spectral density reads
\begin{eqnarray}
	\Sigma_{\beta}(\omega)
	&=&\lim_{T \to \infty}\frac{\mathrm{E}[|{\hat M}_{\beta}^{T}(\omega)|^2]/Z_{\beta}^2}{T}\\
	&=&2\sigma^2
	\mathrm{e}^{-\beta\omega}\sum_{m,n}\mathrm{e}^{-2\beta E_{n}}2\pi
	\delta\left(\omega-(E_{m}-E_{n})\right)|\langle m|M^{x}|n\rangle|^2/Z_{\beta}^2,
	\label{Sigma}
\end{eqnarray}
where
\beq
Z_{\beta}=\mathrm{E}[\langle \Phi{_\beta}| \Phi{_\beta}\rangle].
\eeq

The Fourier transform of the autocorrelation function is given by
\begin{eqnarray}
	G_{\beta}(\omega)=\sum_{m,n}\mathrm{e}^{-\beta E_{n}}2\pi
	\delta\left(\omega-(E_{m}-E_{n})\right)|\langle m|M^{x}|n\rangle|^2/Z_{\beta}.
	\label{Gbetaomega}
\end{eqnarray}
By comparing (\ref{Sigma}) and (\ref{Gbetaomega}),
we obtain a Wiener-Khinchin-like relation for the transverse magnetization:
\begin{eqnarray}
	G_{\beta}(\omega)=2\sigma^2\frac{Z_{\beta/2}^2}{Z_{\beta}}
	\mathrm{e}^{\frac{\beta\omega}{2}}\Sigma_{\beta/2}(\omega).
\end{eqnarray}
The imaginary part of the dynamical susceptibility of interest is given by
\begin{eqnarray}
	\chi''(\omega)=\sigma^2\frac{Z_{\beta/2}^2}{Z_{\beta}}
	\mathrm{sinh}\left(\frac{\beta\omega}{2}\right)\Sigma_{\beta/2}(\omega).
	\label{chiomegaWK}
\end{eqnarray}
Here it should be noted that we need to calculate the quantities of $\beta/2$ (not $\beta)$ to obtain the ESR spectrum of $\beta$.

\begin{figure}[H]
	\begin{center}
		{\includegraphics[width=80mm]{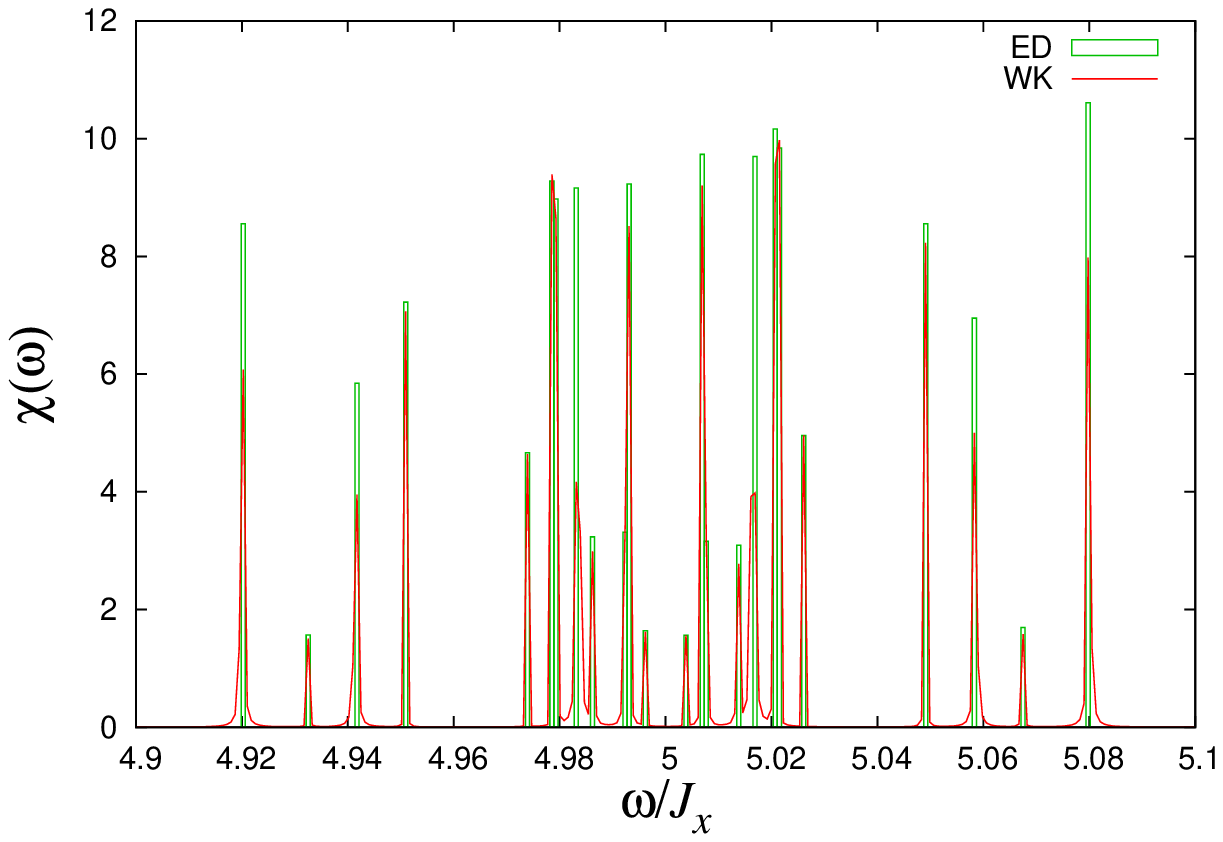}}
		{\includegraphics[width=80mm]{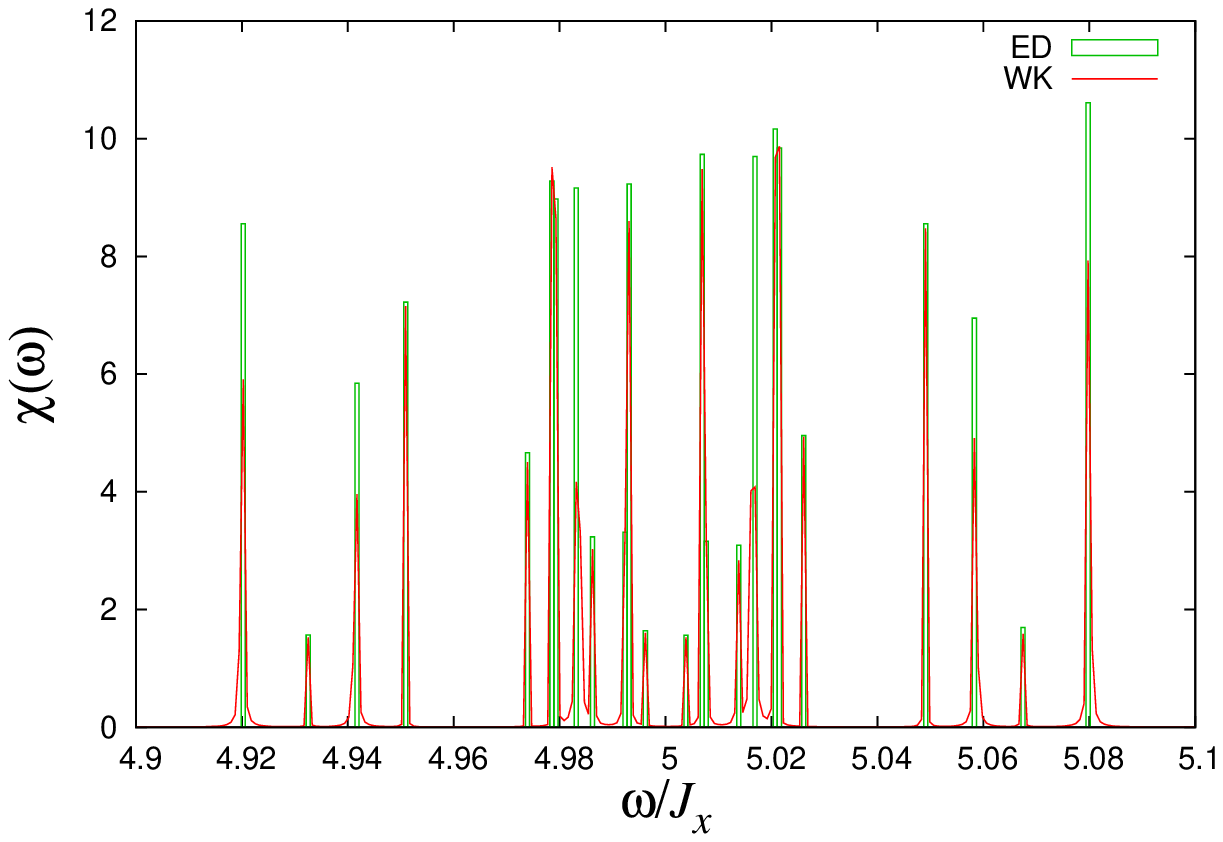}}
	\end{center}
\caption{Absorption spectrum obtained by the ED method (green line) and by the WK method (red line): $N=6$, $\beta^{-1}=100\mathrm{K}$, $H=5\mathrm{K}$,
and the mesh of the frequency $\Delta\omega=2\pi/8192$.
In the ED method (green line), we take the mesh of the frequency $\Delta\omega=2\pi/8192$,
which is the interval of the frequency $2\pi/T$ in the WK method.
(Left) The Wiener-Khinchin spectrum is obtained by calculating Eq.~(\ref{exactWK}) exactly.
(Right) The WK spectrum is obtained by averaging 10000 samples of thermal typical initial states.}
\label{wiener-khinchin}
\end{figure}

In Fig.~\ref{wiener-khinchin}, we show a comparison of the spectrum obtained by the exact diagonalization method and by the WK method.
From Fig.~\ref{wiener-khinchin} it follows that the spectrum is well reproduced without any window procedure,
in contrast to Fig.~\ref{EDTE} (Left) obtained by the AC method.
This fact is attributed to that the finite time effect (Gibbs oscillation) in the WK method gives positive
and smaller artificial peaks in the spectrum (see the Appendix A).

\subsection{Sample distribution of data}\label{sec_variance}

In principle, methods that are based on the thermal typical state require sampling over the random states to obtain data for equilibrium.
However, as mentioned above, the data obtained by the thermal typical state converge to the thermal equilibrium value
as the size of system increases exponentially.
That is, the deviation decreases proportionally to $D^{-1/2}$, where $D$ is the dimension of the Hilbert space.
In AC method, the data are obtained as expectation values of $M^xM^x(t)$, and for large system the convergence to the equilibrium value
is so fast that one initial state suffices to obtain accurate results (see Appendix B for the details).
On the other hand, the WK method extracts information from the quantity ${\hat M}_{\beta}^{T}(\omega)$
whose expectation value is zero, and therefore we have to investigate the fluctuations of the squared variable $|{\hat M}_{\beta}^{T}(\omega)|^2$
because it is not obvious that these fluctuations will decrease rapidly as $D$ increases.

In this section, we study properties of sample distribution of the data.
To characterize the distribution, we study the variance of $|{\hat M}_{\beta}^{T}(\omega)|^2$:
\begin{equation}
\mathrm{Var}\left(|{\hat M}_{\beta}^{T}(\omega)|^2\right)
=\mathrm{E}\left[
|{\hat M}_{\beta}^{T}(\omega)|^4 \right]
- \mathrm{E}\left[|{\hat M}_{\beta}^{T}(\omega)|^2\right]^2.
\label{MX4text}
\end{equation}
To this end, we introduce
\begin{eqnarray}
{\widetilde M}_{\beta}^{T}(\omega)&\equiv& \langle\Phi|A(\omega)|\Phi\rangle,
 \\
\noalign{\noindent where}
A(\omega)&=&\mathrm{e}^{-\beta \mathcal{H}/2}Y(\omega) \mathrm{e}^{-\beta \mathcal{H}/2},
 \\
\noalign{\noindent and}
Y(\omega)&=& \frac{1}{T}\int_{-T/2}^{T/2} \mathrm{e}^{-i\omega t} \mathrm{e}^{\mathrm{i}\mathcal{H}t} M^{x} \mathrm{e}^{-\mathrm{i}\mathcal{H}t}\mathrm{d}t.
\label{MX0text}
\end{eqnarray}
which is proportional to Eq.~(\ref{mxbetaomega}), i.e.,
${\widetilde M}_{\beta}^{T}(\omega)={\hat M}_{\beta}^{T}(\omega)\times \mathrm{e}^{i\omega T/2}/T$.
The reason for shifting the interval from $[0,T]$ to $[-T/2,T/2]$ is to assure
that $Y(\omega)=Y^\dagger(-\omega)$ and
\begin{eqnarray}
A_{ap}^{\phantom{\ast}}\equiv A(\omega)_{ap}=\left(A(-\omega)_{pa}\right)^{\ast}
=\left(A^\dagger(\omega)_{pa}\right)^{\ast}
=\frac{\sin\left[(\omega-(E_{a}-E_{p}))T/2\right]}{[\omega-(E_{a}-E_{p})]T/2}
\langle a|\mathrm{e}^{-\beta \mathcal{H}/2}M^{x}\mathrm{e}^{-\beta \mathcal{H}/2}|p\rangle,
\label{MX1text}
\end{eqnarray}
where we choose the states $|a\rangle$ to be the eigenstates of $\mathcal{H}$.
With this choice, $\langle a|M^{x}|a\rangle=0$ and $A_{aa}=0$ which simplifies the calculations significantly.
Using these quantities,
\begin{equation}
|{\widetilde M}_{\beta}^{T}(\omega)|^2
=
\sum_{a,b,p,q}\xi_{a}^{\ast}\xi_{b}^{\ast} \xi_{p}^{\phantom{\ast}}\xi_{q}^{\phantom{\ast}} A_{ap}A_{qb}^\ast.
\label{MX2text}
\end{equation}
and, using the identities Eq.~(\ref{RS30text}), the expectation of Eq.~(\ref{MX2text}) is expressed as
\begin{equation}
\mathrm{E}[|{\widetilde M}_{\beta}^{T}(\omega)|^2]= \sum_{a,b}A_{ab}^{\phantom{\ast}}A_{ab}^\ast
= 4\sigma^4\mathbf{Tr\,} A A^\dagger.
\label{MX3text}
\end{equation}
Hereafter, for simplicity of notation but without loss of generality, we choose $2\sigma^2=1$.
Note that in Eq.~(\ref{MX3text}), we did not yet include other terms in (\ref{chiomegaWK}), i.e., $Z_{\beta}$.
But, as we mentioned above, the variances of them are small and thus
we only concern the variance (\ref{MX4text}).

We calculate the first term in Eq.~(\ref{MX4text}) by making extensive use of the properties
of Gaussian random variables.
We have
\begin{eqnarray}
\mathrm{E}\left[
|{\widetilde M}_{\beta}^{T}(\omega)|^4 \right]
&=&
\sum_{a,b,c,d}
\sum_{p,q,r,s}
\mathrm{E}\left[
\xi_{a}^{\ast}\xi_{b}^{\ast}\xi_{c}^{\ast}\xi_{d}^{\ast}
\xi_{p}^{\phantom{\ast}}\xi_{q}^{\phantom{\ast}}\xi_{r}^{\phantom{\ast}}\xi_{s}^{\phantom{\ast}}
\right]
A_{ap}^{\phantom{\ast}}A_{qb}^\ast
A_{cr}^{\phantom{\ast}} A_{s d}^{\ast}
\nonumber \\
&=&
\sum_{a,b,c,d}
\sum_{p,q,r,s}
\bigg\{
\mathrm{E}\left[
\xi_{a}^{\ast}
\xi_{p}^{\phantom{\ast}}
\right]
\mathrm{E}\left[
\xi_{b}^{\ast}\xi_{c}^{\ast}\xi_{d}^{\ast}
\xi_{q}^{\phantom{\ast}}\xi_{r}^{\phantom{\ast}}\xi_{s}^{\phantom{\ast}}
\right]
+
\mathrm{E}\left[
\xi_{a}^{\ast}
\xi_{q}^{\phantom{\ast}}
\right]
\mathrm{E}\left[
\xi_{b}^{\ast}\xi_{c}^{\ast}\xi_{d}^{\ast}
\xi_{p}^{\phantom{\ast}}
\xi_{r}^{\phantom{\ast}}
\xi_{s}^{\phantom{\ast}}
\right]
\nonumber \\
&&\hbox to 2cm{}
+
\mathrm{E}\left[
\xi_{a}^{\ast}
\xi_{r}
\right]
\mathrm{E}\left[
\xi_{b}^{\ast}\xi_{c}^{\ast}\xi_{d}^{\ast}
\xi_{p}^{\phantom{\ast}}
\xi_{q}^{\phantom{\ast}}
\xi_{s}^{\phantom{\ast}}
\right]
\bigg.
+
\mathrm{E}\left[
\xi_{a}^{\ast}
\xi_{s}^{\phantom{\ast}}
\right]
\mathrm{E}\left[
\xi_{b}^{\ast}\xi_{c}^{\ast}\xi_{d}^{\ast}
\xi_{p}^{\phantom{\ast}}
\xi_{q}^{\phantom{\ast}}
\xi_{r}^{\phantom{\ast}}
\right]
\bigg\}
A_{ap}^{\phantom{\ast}}A_{qb}^\ast
A_{cr}^{\phantom{\ast}} A_{s d}^{\ast}
.
\label{MX5text}
\end{eqnarray}
The first term in the curly brackets vanishes because
$\mathrm{E}\left[\xi_{a}^{\ast}\xi_{p}\right]=\delta_{a,p}$ and $A_{aa}=0$.
Interchanging the summation indices $(r,q)$ and $(s,q)$ in the third
and fourth term, respectively we obtain
\begin{eqnarray}
\mathrm{E}\left[
|{\widetilde M}_{\beta}^{T}(\omega)|^4 \right]
=
4\mathbf{Tr\,} AA A^\dagger A^\dagger
+\left|\mathbf{Tr\,} A^2 \right|^2
+2\left(\mathbf{Tr\,} AA^\dagger\right)^2
+2\mathbf{Tr\,} AA^\dagger A A^\dagger
,
\label{MX6text}
\end{eqnarray}
and using Eq.~(\ref{MX3text}) we finally obtain for the variances
\begin{eqnarray}
\mathrm{Var}\left(|{\widetilde M}_{\beta}^{T}(\omega)|^2\right)
&=&
4\mathbf{Tr\,} AA A^\dagger A^\dagger
+\left|\mathbf{Tr\,} A^2 \right|^2
+\left(\mathbf{Tr\,} AA^\dagger\right)^2
+2\mathbf{Tr\,} AA^\dagger A A^\dagger.
\label{MX7ztext}
\end{eqnarray}

We want to find bounds to the estimate of the variance Eq.~(\ref{MX7ztext}).
Let us denote the (non-negative) eigenvalues of $AA^\dagger$ ($A^\dagger A$) by
$x_k^2$ ($y_k^2=x_k^2$), that is
$AA^\dagger|x_k\rangle=x_k^2|x_k\rangle$
($A^\dagger A|y_k\rangle=y_k^2|y_k\rangle$).
We assume that the eigenvectors of $AA^\dagger$ and $A^\dagger A$ are normalized.
Then we have
\begin{eqnarray}
\left(\mathbf{Tr\,} AA^\dagger\right)^2&=&
\left(\sum_k x_k^2\right)^2\ge \sum_k x_k^4=\mathbf{Tr\,} AA^\dagger A A^\dagger,\\
\mathbf{Tr\,} AA^\dagger A^\dagger A &=&
\sum_{k,l} x_k^2 y_l^2 |\langle x_k|y_l\rangle|^2
\le \sum_{k,l} x_k^2 y_l^2 = \left(\mathbf{Tr\,} AA^\dagger\right)\left(\mathbf{Tr\,} A^\dagger A\right)
= \left(\mathbf{Tr\,} AA^\dagger\right)^2.
\label{MX10text}
\end{eqnarray}
As $(X,Y)\equiv \mathbf{Tr\,} X^\dagger Y$ defines a scalar product,
by the Schwarz inequality
$|(A^\dagger,A)|^2=\left|\mathbf{Tr\,} A^2 \right|^2\le \left|\mathbf{Tr\,} AA^\dagger \right|^2$.
Noting that the last term in Eq.~(\ref{MX7ztext}) cannot be negative, putting all this together we find
\begin{eqnarray}
1\le
1+\frac{
4\mathbf{Tr\,} AA A^\dagger A^\dagger
+\left|\mathbf{Tr\,} A^2 \right|^2
+2\mathbf{Tr\,} AA^\dagger A A^\dagger}{\left(\mathbf{Tr\,} AA^\dagger\right)^2}
&\le&8
.
\label{MX7btext}
\end{eqnarray}
Thus, we find the variance is bounded as
\begin{eqnarray}
1<
{\mathrm{Var}\left(|{\hat M}_{\beta}^{T}(\omega)|^2\right)
\over
\mathrm{E}\left[|{\hat M}_{\beta}^{T}(\omega)|^2\right]^2}
<8.
\label{MX70text}
\end{eqnarray}
indicating that in the WK method, it is necessary to average over different initial states, also for large systems.
This is in sharp contrast to the case of the AC method in which the variance vanishes exponentially with the system size.
But Eq.~(\ref{MX70text}) also indicates that the variance is bounded, independent of the system size, and therefore
we can obtain good estimates for the mean from finite samples.

We emphasize that the variance of the WK method does not depend on $D$, that is, the distribution shows a kind of typicality.
Thus we can obtain the correct mean with aimed precision by sampling, regardless of the system size.


\section{Distribution of average value and convergence in distribution}\label{sec_dist}

In this section we investigate the sample distribution of $\chi(\omega)$ data.
For concreteness, we choose the value of $\omega$ to be 4.9916$\mathrm{K}$ at which the spectrum has a maximum (peak).

\begin{figure}[H]
	\begin{center}
	\includegraphics[width=80mm]{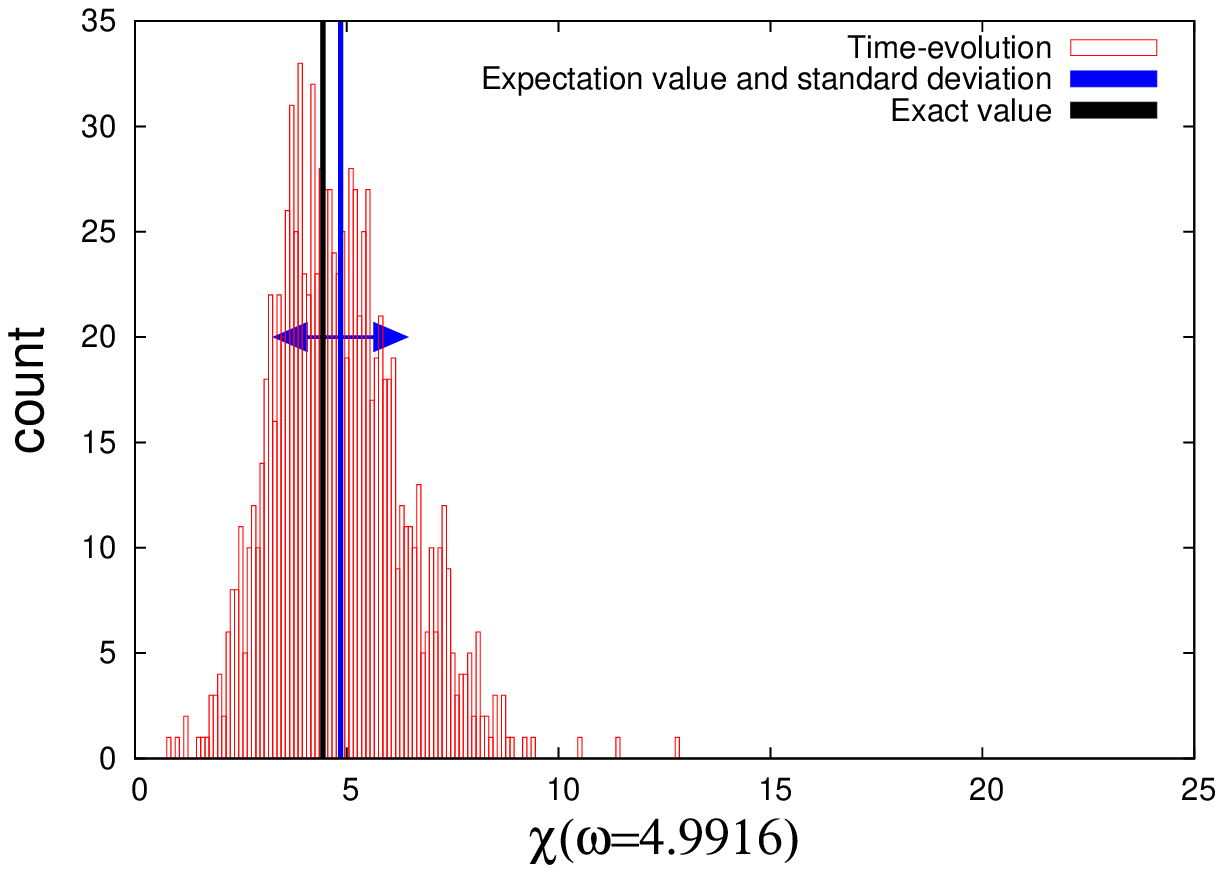}
	\includegraphics[width=80mm]{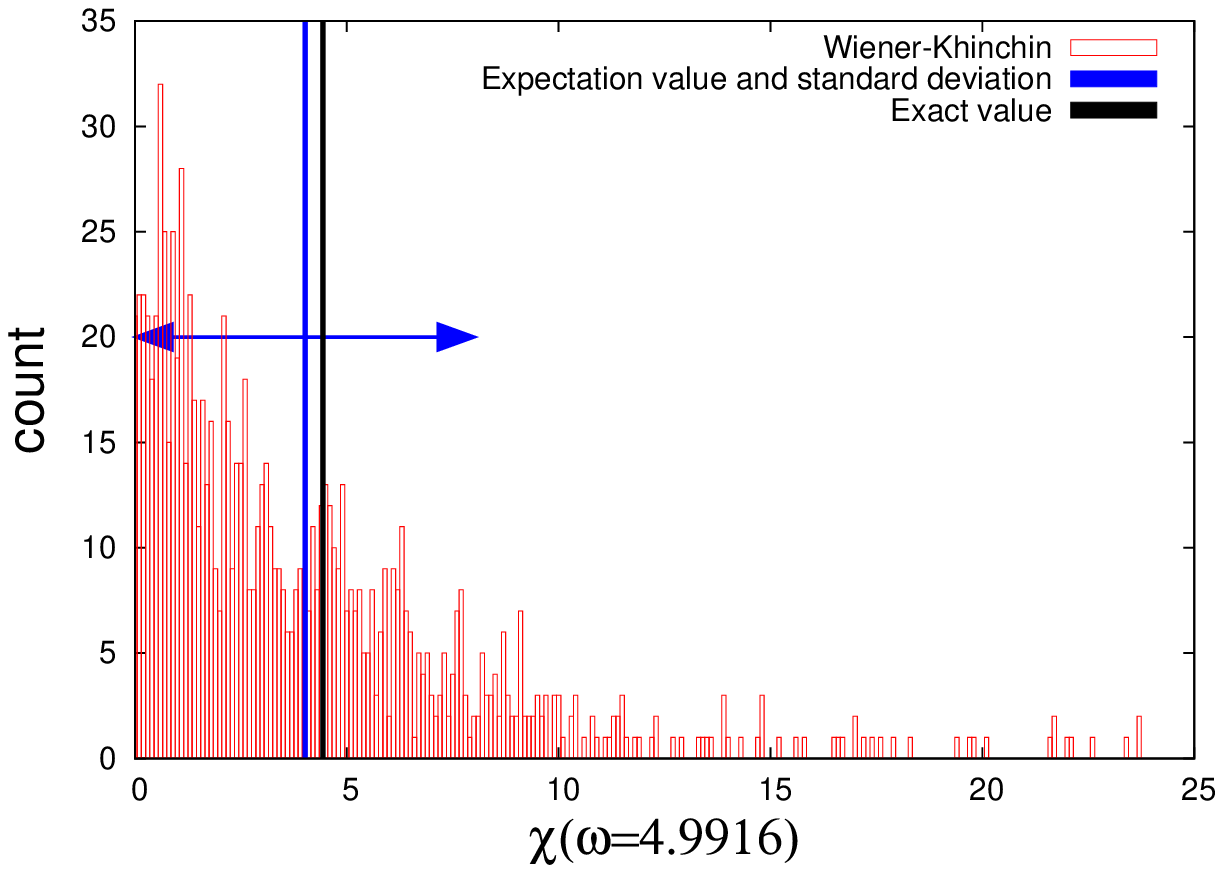}
	\end{center}
	\caption{(Left)
	The distribution of values over 1000 samples obtained by the AC method at $\omega=4.9916\mathrm{K}$ for $N=10$.
The mean (=4.8508$\mathrm{K}$) is given by the blue line, and the standard deviation (=1.5049$\mathrm{K}$) is shown by the arrow.
The exact value is denoted by the black solid line whose $\omega=4.4666\mathrm{K}$.
(Right) The distribution of values over 1000 samples obtained by the WK method at $\omega=4.9916\mathrm{K}$ for $N=10$.
The mean (=4.01579$\mathrm{K}$) is given by the blue line, and the standard deviation (=3.9749$\mathrm{K}$) is shown by the arrow.
The exact value is denoted by the black solid line whose $\omega=4.4666\mathrm{K}$.
}
	\label{N10AC_histogram}
\end{figure}

First we show the distribution of data in the AC method for the system $N=10$ where $D=1024$ and $D^{-1}\approx0.001$ is rather small.
The histogram of this data is given in Fig.~\ref{N10AC_histogram}(Left).
The exact value is denoted by the black line. The mean of the distribution is shown by the bold blue line.
The standard deviation (square-root of the variance) is given by the arrow.
Clearly Fig.~\ref{N10AC_histogram}(Left) shows that the data is distributed around the correct value with RSD=0.31025$\mathrm{K}$.
As we already menioned several times, as the size of the system increases the variance will vanish as $D^{-1}$.

The histogram obtained by the WK method is depicted in Fig.~\ref{N10AC_histogram}(Right).
In contrast to the AC method, the distribution is of the exponential type.
From the figure, we find the variance is about the same as the average, i.e., RSD=0.98982$\mathrm{K}$.
This value corresponds to the lower bound of the estimation (\ref{MX70text}).
This fact will be discussed in more detail below.

We present the spectra for $N=10$ obtained by the both methods compared with the exact one obtained by the ED method.
Because we have the exact eigenstates and eigenvalues,
we computed the results of AC and WK methods by employing Eqs.~(\ref{mxbetaomega}) and (\ref{Gbetaomega})
using the exact eigenstates and eigenvalues (not by sampling).
\begin{figure}[H]
	\begin{center}
	\includegraphics[width=80mm]{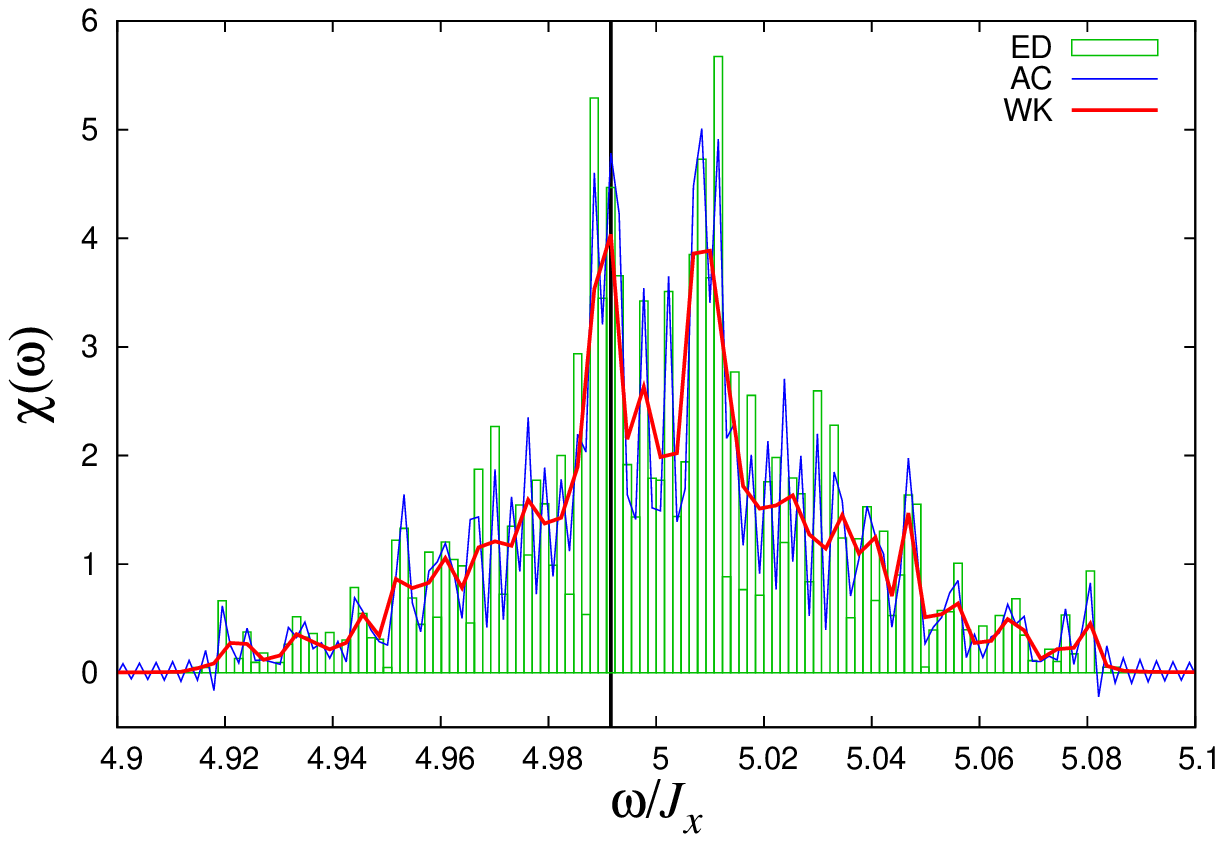}
	\includegraphics[width=80mm]{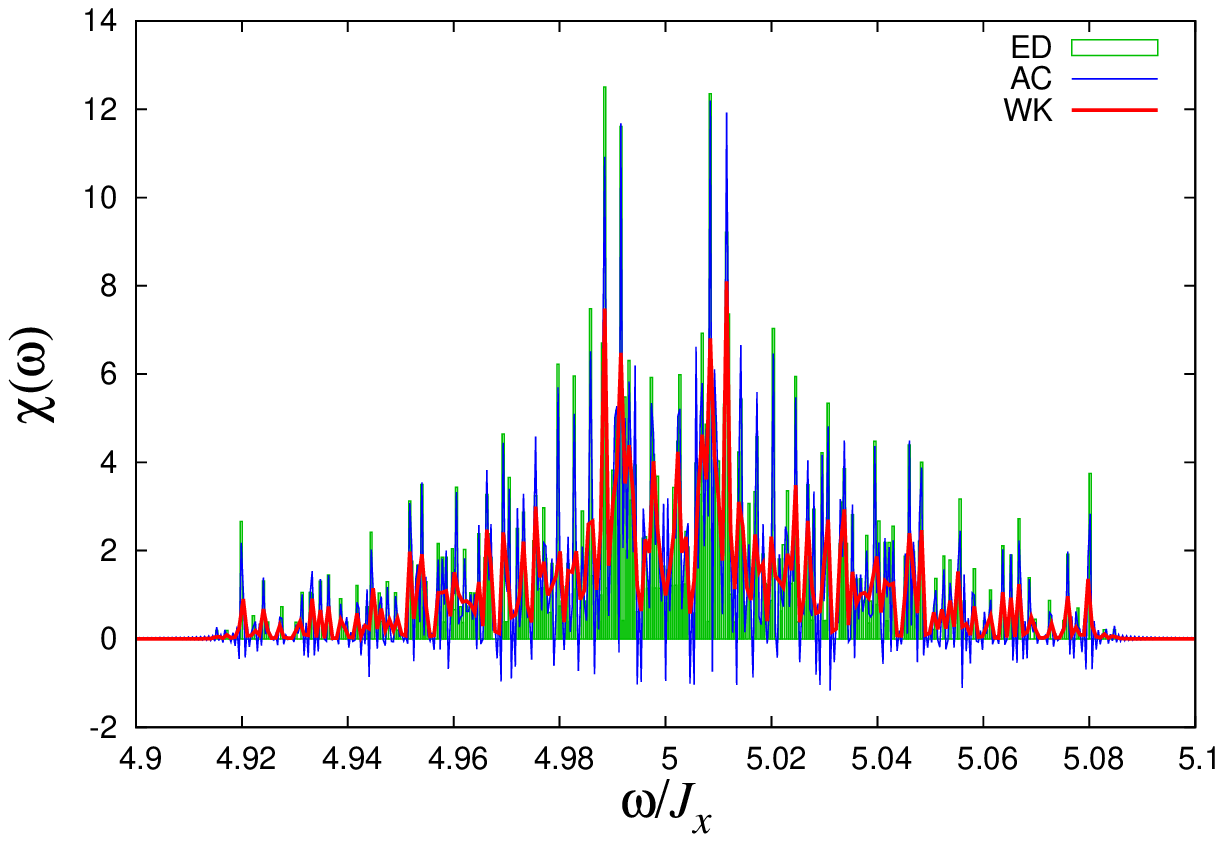}
	\end{center}
\caption{
(Left) Spectra obtained by the AC method (blue line) by the WK method (red line) with
$n_t=4096$. The column histogram is obtained by the ED method.
The mesh of the frequency $2\pi/4096$ for the WK method and $\pi/4096$ for the AC method and the ED method.
The line $\omega=4.9916\mathrm{K}$ is denoted by the black line.
(Right) $n_t=16384$, and the mesh of the frequency $2\pi/16384$ for the WK method and $\pi/16384$ for the AC method and the ED method.}
\label{N10_exact}
\end{figure}
Although both methods approximately reproduce the exact results,
the AC method shows large fluctuations for both $n_t$ large and small.
On the other hand, the WK method gives rather good agreement.

In the case of the AC method, the obtained values have a narrow distribution around the mean,
and it is known that the variance decreases as $D^{-1}$.
In this sense, 
the AC method
 has an advantage.
But the WK method also gives correct estimates by sampling and the obtained shape of the spectrum is close to the exact values.

\subsection{Special case 1: Energy gap selection}
In the previous section, we found that the relative variance is close to one, the minimum value of the bound Eq.~(\ref{MX70text}).
In this subsection, we consider the reason of this fact.
To derive Eq.~(\ref{MX70text}), we considered the general case.
But we can also use the fact that $A_{ap}$ (Eq.~(\ref{MX1text})) is almost zero except when $\omega\simeq E_a-E_p$.
If we assume that there is a unique set $(n,l,m)$
which satisfies the conditions $E_m-E_n=E_m-E_l=\omega$ and $\langle m|M^x|n\rangle\langle m|M^x|l\rangle\ne 0$ then
all terms are negligibly small except for the term $\left(\mathbf{Tr\;}AA^{\dagger}\right)^2$.
Therefore
\beq
	\frac{\sqrt{\mathrm{Var}(|{\hat M}_{\beta}^{T}(\omega)|^2)}}{\mathrm{E}[|{\hat M}_{\beta}^{T}(\omega)|^2]}\simeq {\left(\mathbf{Tr\;}AA^{\dagger}\right)^2
       \over \left(\mathbf{Tr\;}AA^{\dagger}\right)^2}=1,
\eeq
which is indeed what we have observed in our actual calculations.

\subsection{Special case 2: Isolated delta function}

We give the variance of the WK method for the case where
the spectrum consists of well-separated delta functions which are given only by a single resonant pair of states, that is
the non-resonance condition that if $E_m-E_n=E_{m'}-E_{n'}$ then ($m=m'$ and $n=n'$) or ($m=n$ and $m'=n'$) is satisfied.
Then, we can use the fact that approximately $\delta^T(x)=\delta(x)$ and we have
\begin{eqnarray}
	\mathrm{E}[|{\hat M}_{\beta}^{T}(\omega)|^2]
	=2\sigma^2T\mathrm{e}^{-\beta\omega}\times
	\mathrm{e}^{-2\beta E_{n}}2\pi
	\delta^{T}\left(\omega-(E_{m}-E_{n})\right)|\langle m|M^{x}|n\rangle|^2.
\end{eqnarray}
The variance is estimated by making use of the relation:
\beq
\mathrm{E}\left[\xi_m^*\xi_n\xi_{m'}\xi_{n'}^*\xi_k^*\xi_l\xi_{k'}\xi_{l'}^*\right]
=\mathrm{E}\left[(\xi_m^*\xi_m)^2(\xi_{n}^*\xi_{n})^2\right]
\simeq 4
\eeq
as
\beq
\mathrm{E}\left[
|{\hat M}_{\beta}^{T}(\omega)|^4 - \mathrm{E}[|{\hat M}_{\beta}^{T}(\omega)|^2]^2
\right]\simeq 3\mathrm{E}[|{\hat M}_{\beta}^{T}(\omega)|^2]^2.
\eeq
Thus, the relative variance is 3, a result which does not depend on $D$.

In the AC method, the variance of sampled amplitude of the isolated delta functions is the same
as the expectation value for the present case and ${\rm RSD} \ge 1$.
However, the isolated delta function in the spectrum is only relevant in small systems.
For large system with large $D$, the amplitude of
the isolated delta function is small and we can ignore it.

\section{The total amplitude of the spectrum} 
\label{amplitude}
\begin{figure}[H]
	\begin{center}
	\includegraphics[width=100mm]{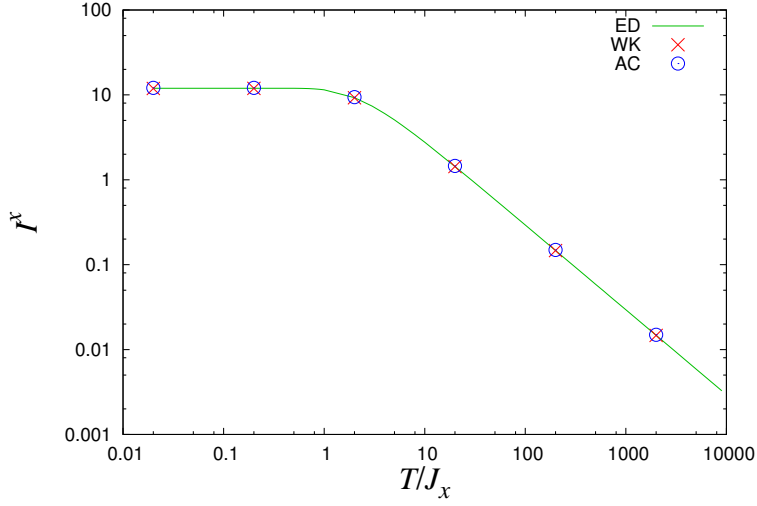}
	\end{center}
	\caption{Comparison of the intensities of the absorption as a function of the temperature
obtained by the WK method (red cross)  and by the AC method (blue open circle). The green solid curve is the exact value: $N=6$, $H=5\mathrm{K}$.}
\label{intensity}
\end{figure}

The total amplitude of the spectrum, or the intensity, i.e. the integral over the absorption spectrum (Eq.~(\ref{Ixomega})), is given by
\beq
I^x=\int_0^{\infty} I^x(\omega)\mathrm{d}\omega.
\eeq
It is one of the fundamental pieces of information obtained in ESR experiments.
Its temperature dependence has been calculated for the single molecular magnet V$_{15}$\cite{machida}.
In this subsection, we demonstrate that $I^x$ obtained by the WK method correctly reproduces the exact ED results.
Details of the analysis are given in Appendix A.
In the WK method, a delta function is represented by the sinc-function
\beq
\delta(\omega) \rightarrow {1\over T}\left( {\sin\omega T/2\over \pi\omega}\right)^2.
\label{WKdelta}
\eeq
Because of the relation
\beq
{1\over T}\int_{-\infty}^{\infty}\left( {\sin\omega T/2\over \pi\omega}\right)^2\mathrm{d}\omega=\frac{1}{2\pi},
\label{H2}
\eeq
the intensity $I^x$ is obtained correctly by analytical integration.
However, when the DFT yields the spectrum at discrete points $\omega=2k\pi/T,\quad k=0,1,2,\cdots, n_t-1$ only.
Thus, the integral over the $\omega$ is given by a discrete sum.
In Appendix A, we demonstrate that the discrete sum agrees with the analytical result.
We also show that for the AC method in which the discrete points are given by $\omega=k\pi/T,\quad k=-n_t,\cdots,-1,0,1,2,\cdots, n_t-1$
we recover the  analytical result of the intensity, in spite of Gibbs oscillations.
We calculate the results of AC and WK methods by evaluating the formulae for the expectation value
by using the exact eigenvalues and eigenstate, as we did for  Fig.~\ref{EDTE} and \ref{wiener-khinchin}(Left),
i.e., without sampling over thermal typical states.
In Fig.~\ref{intensity}, we show that both the AC and WK methods reproduce the results of the ED method.


\section{Applications to large systems}\label{sec_APPLICATION}

As an example of application of the methods to larger systems, we compute the spectrum by the AC and WK method for $N=16$ with $n_t=$16384.
For this size, we can still compute all eigenvalues and eigenstates, hence we know the positions of the eigenfrequencies and the corresponding amplitudes.
The ensemble of delta functions is turned into a histogram by using a frequency mesh $\Delta\omega=2\pi/T$, ($T=16384\times \mathrm{d}t(=0.5))$.

\begin{figure}[H]
	\begin{center}
	\includegraphics[width=80mm]{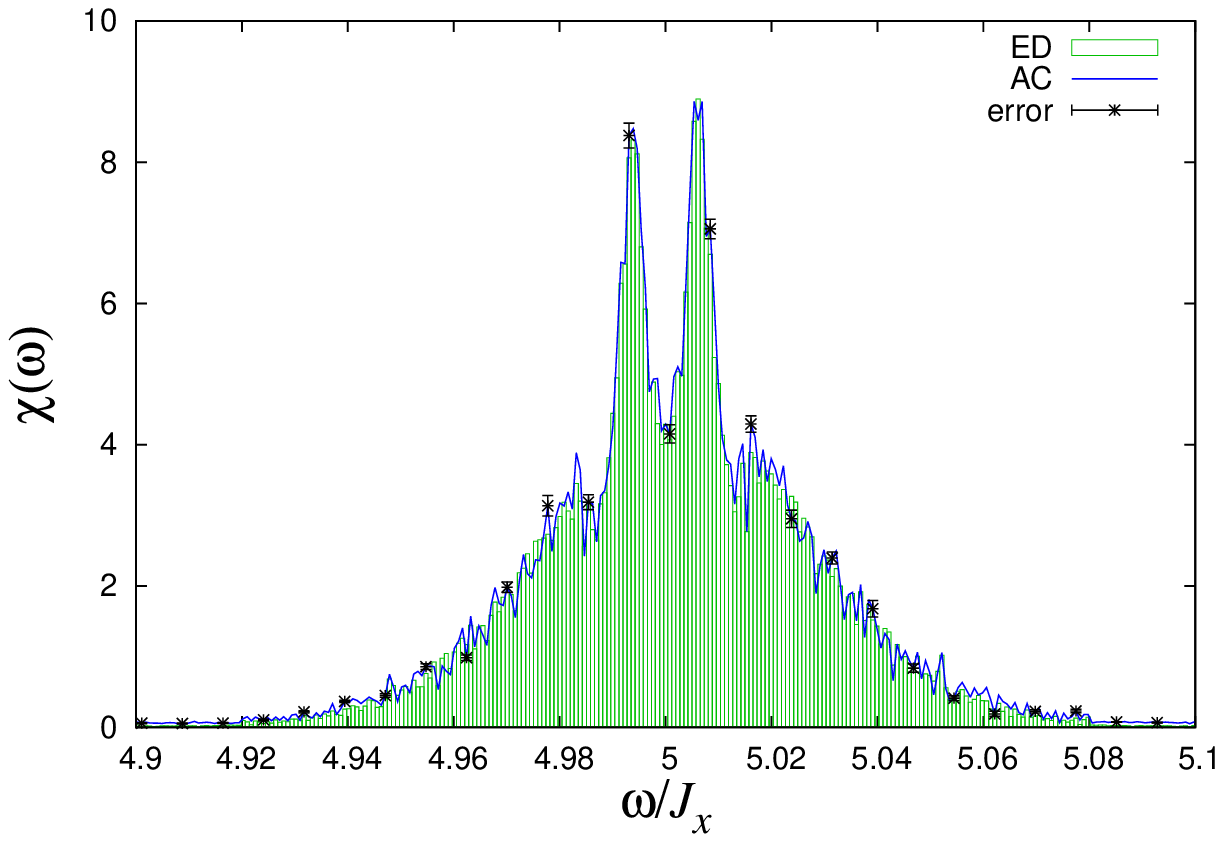}
	\includegraphics[width=80mm]{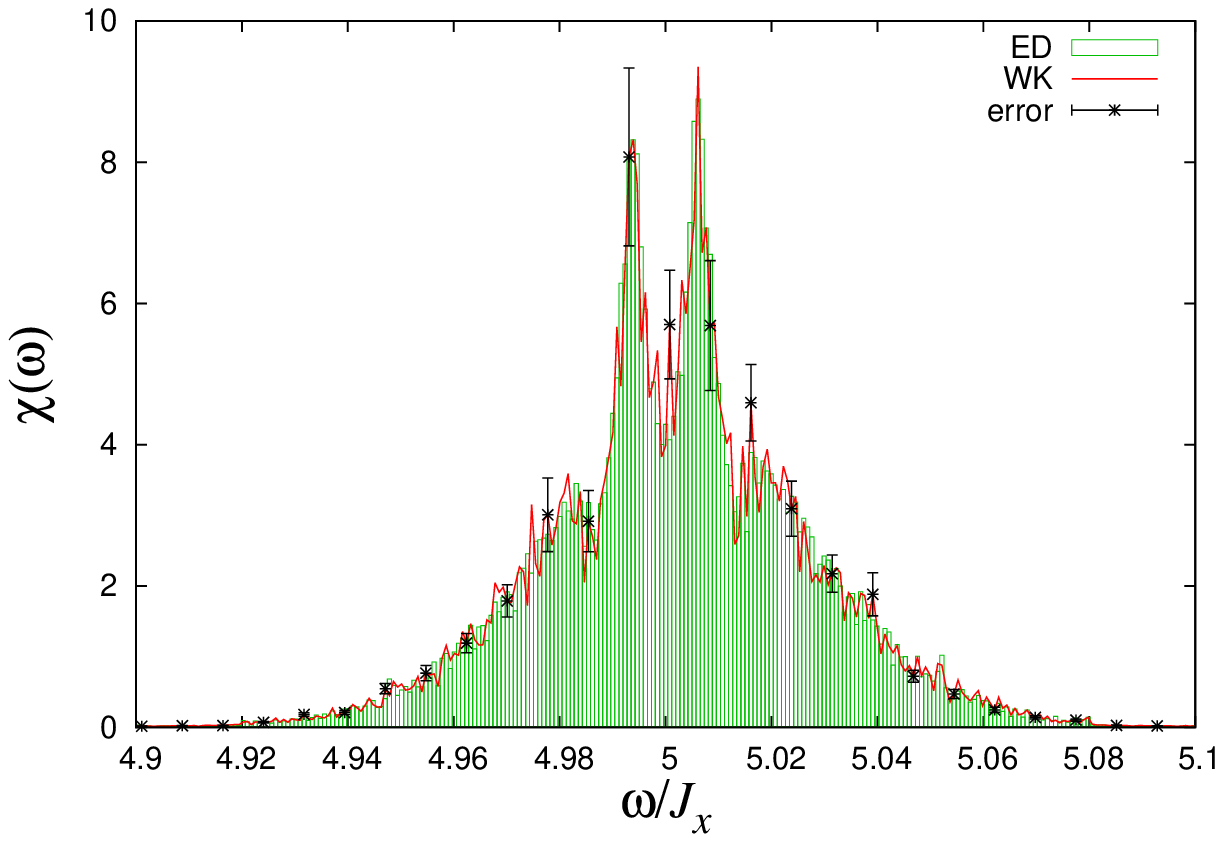}
	\end{center}
	\caption{
(Left) Spectrum for $N=16$. The histogram is obtained by exact diagonalization,
points with error bars are the data obtained by the AC method with 10 samples and the time interval $[0,T=16384\times \mathrm{d}t], \mathrm{d}t=0.5$.
(Right) Results of the WK method with 52 samples over the sme time interval
}
	\label{N16ACED}
\end{figure}

In Fig.~\ref{N16ACED}(Left),
 we present a comparison between the spectra obtained by ED and AC methods.
The data obtained by the AC method shows sharp features because we did not use a window in the DFT procedure.
As the statistical fluctuations on the autocorrelation are small, 10 samples suffice to find almost perfect agreement with the ED result.
Therefore, we may 
conclude
 that as the size of the system increases, the effects of finiteness of the time interval
reflected in e.g. Gibbs oscillations seems to be small whereas we found serious effects in the case $N=6,10$.
In Fig.~\ref{N16ACED}(Right), we show the comparison between the spectra obtained by the ED and WK method.
In the WK method, we took 52 samples, and we did not use a window function for the DFT procedure.
Here we again find a good agreement but the variance of ensemble average is about the same as the mean as we found in the case $N=10$.
We conclude that for $N=16$, the results of both methods are in good agreement those of ED.

\begin{figure}[H]
	\begin{center}
	\includegraphics[width=59mm]{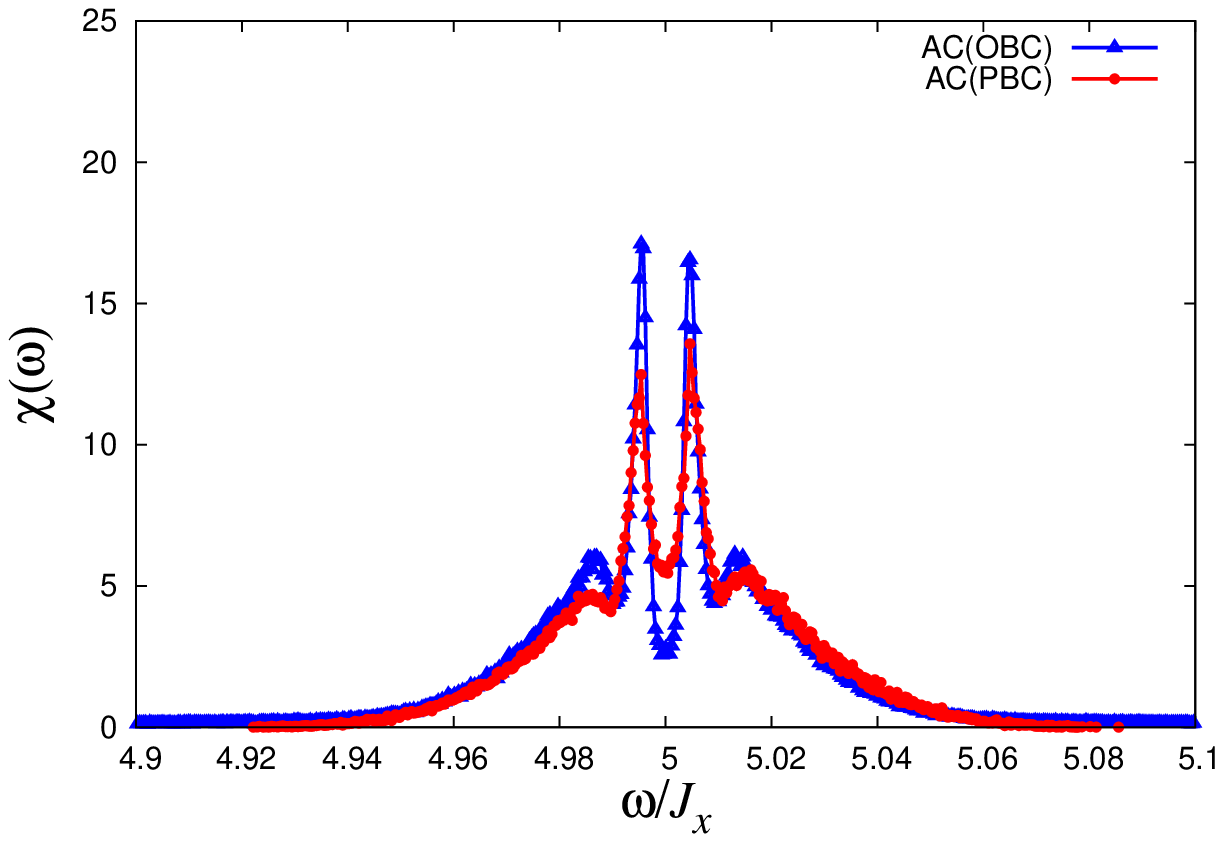}
	\includegraphics[width=59mm]{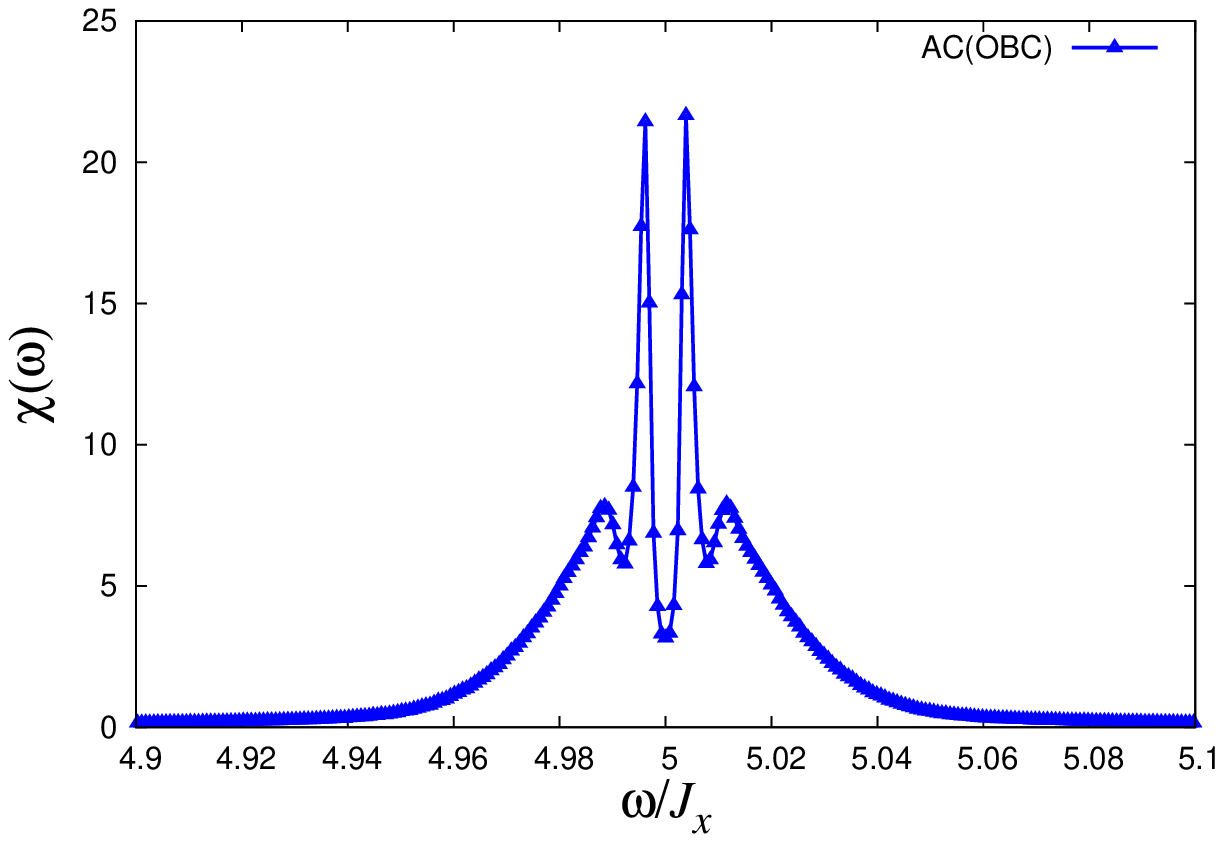}
	\includegraphics[width=59mm]{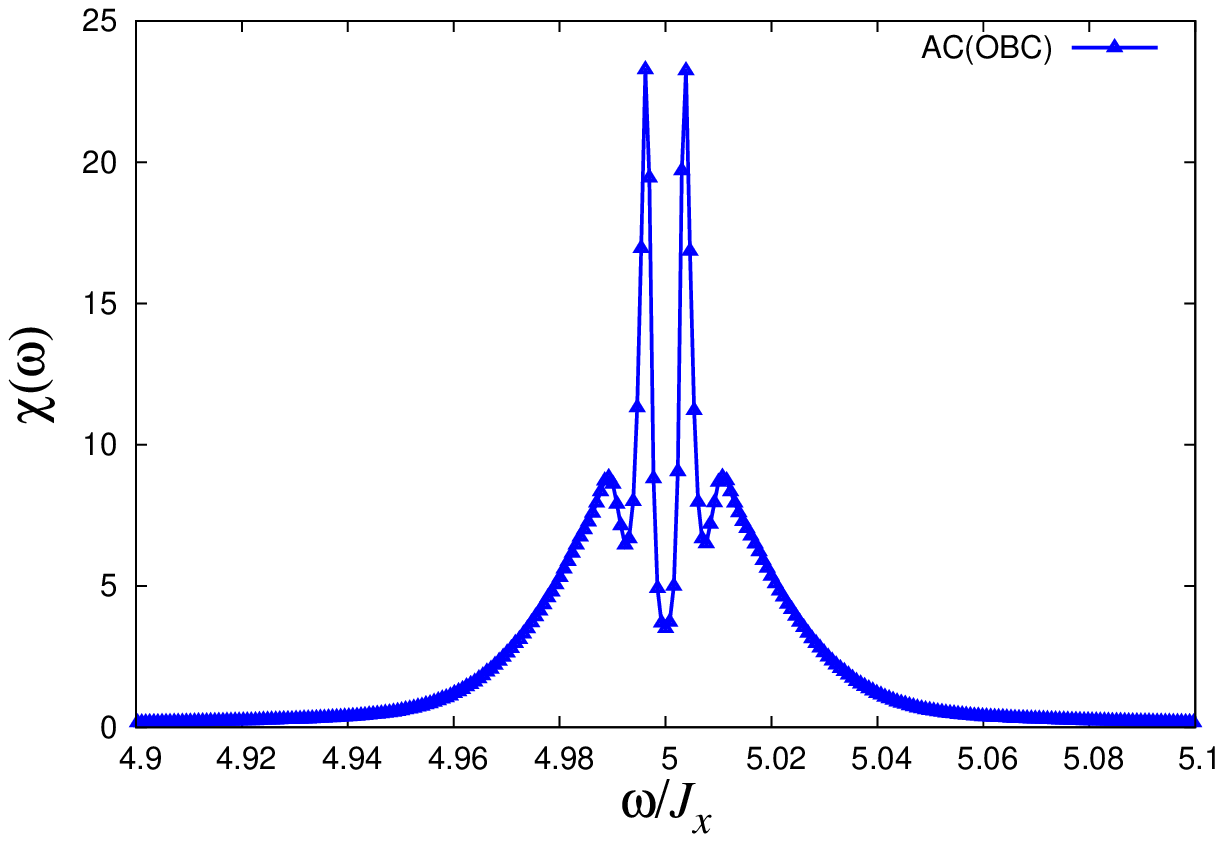}
	\end{center}
	\caption{Spectrum obtained by the AC method with the open boundary condition (blue line). (Left) $N=20$. The red line denotes the spectrum with the periodic boundary condition. (Center) $N=24$. (Right) $N=26$.}
	\label{N20AC}
\end{figure}

With above observation, we also obtained spectra for $N\ge 20$ by the AC method.
Up to now, we adopted the periodic boundary conditions (PBC). For a change and also because open boundary condition (OBC)
is of interest because they occur in actual materials, we show results for $N=20$ obtained for both boundary conditions.
In Fig.~\ref{N20AC}(Left) , the red curve (circles) is the spectrum obtained with PBC
and the blue curve with triangles is that of OBC.
We find that the spectrum of OBC has a shaper double peaks in the center,
but the global shape is similar in both cases.
Finally, we present spectra for $N=24,26$ for OBC in Fig.~\ref{N20AC} (Center) and (Right), respectively.
We used $n_t=16384$ with $\mathrm{d}t=0.5$ for $N=20$, and $n_t=8192$ with $\mathrm{d}t=0.5$ for $N=24,26$ and did not use any windowing procedure.
From the size dependence, we conclude that the double peak structure will survive in the large size limit.

\section{Summary and Discussion}\label{sec_summary}  

We have studied time-domain methods to compute the ESR spectrum through sampling of thermal typical states.
Although it has been known that the autocorrelation function can be obtained with a few samples
thanks to the typicality property, practical aspects of these calculations have not yet been explored.

We proposed a method to obtain the spectrum from the motion of $M^x$ by making use of the Wiener-Khinchin-like relation.
The dynamics of magnetization from sampled initial states provides an estimate the spectrum density.
But in the quantum equilibrium state the expectations value of $M^x(t)$ and the sampled data for the spectrum density $M^x_{\beta}(\omega)$ are zero.
Then, we take thermal typical states as initial states
,and obtain the average of squared spectrum density
 $|M^x_{\beta}(\omega)|^2$,
i.e., the Fourier transform of the autocorrelation function.
To this end, we proposed a quantum mechanical version of the Wiener-Khinchin relation.
However, it should be mentioned that the WK method does not directly relate to the experimental situation.
Because $\langle M^x_{\beta}(\omega)\rangle_{\rm eq}=0$, the averaged squared spectrum density
$|M^x_{\beta}(\omega)|^2$
does not converge to a mean value but has a finite distribution, also for large systems.
Therefore the convergence property of the 
thermal
typical state does not apply in this case.
We studied statistical properties of the distribution
and found that form of the distribution converges in the sense of typicality.
We derived bounds on the relative variance of the distribution.

In time domain methods the effect of finite observation time is an important issue.
We studied this aspect in detail for both the AC and WK method.
It was found that for small systems, the AC method is suffering from severe effect of finiteness of $T$ while
in WK method the effect is practically suppressed.
The WK and AC methods give complementary information for the effects of finite $T$
and a comparison between them is useful to confirm that the obtained result is not affected by the finiteness observation time.
We also found that for both the AC and WK method, the effect of finiteness of $T$
decreases
 with increasing system size.
As the system size increases, the efficiency of the AC method increases too.

We presented the spectrum for a one-dimensional XXZ chain up with to $N=26$ spins,
where the spectrum shows a double peak which will be sustained in the thermodynamic limit.
In our numerical work, we focussed on chains with an even number of spins.
The ESR spectra of chains with an odd number of spins show behavior that is qualitatively different from
the one of chains with an even number of spins, meriting a study in it own right.
As the focus of the present paper was on time-domain methods rather than on specific applications,
we relegate this study to future research.

In summary, in this paper, we proposed the new method to compute the ESR spectrum by making use of mathematical relations discussed
in Sec \ref{sec_WK}. We have studied one particular time-domain method based on the Wiener-Khinchin theorem
but there are many other ways to obtain the spectrum density by making use of relations similar to Wiener-Khinchin theorem.
For example, if we calculate ${\rm E}[\langle\Phi_{\beta}|M^x|\Phi_{\beta}\rangle\langle\Phi_{\beta}|M^x(t)|\Phi_{\beta}\rangle]$,
we can obtain a similar expression. Studying the relation among them is an interesting problem for future research.

\section*{Acknowledgements}

The present work was supported by
Grants-in-Aid for Scientific Research C (25400391) from MEXT of Japan, and the Elements
Strategy Initiative Center for Magnetic Materials under the
outsourcing project of MEXT. The numerical calculations were supported by the supercomputer center of ISSP of Tokyo University.
We also acknowledge the JSPS Core-to-Core Program: Non-equilibrium dynamics of soft matter
and information.

\appendix

\section{Effect of finiteness of the observation time domain and window function}
\label{sec:appendixA}

\subsection{Effect of Gibbs oscillation}
Because the observation time is finite $[-T, T]$ and we have a finite number of points in time.
the discrete Fourier transform (DFT) gives amplitude at the discrete points
\beq
\omega_k=\Delta\omega\times k, \quad \Delta\omega={\pi\over T},\quad k:{\rm integer}.
\eeq
On the other hand, the resonance points are not necessarily at those points.
If the spectrum has a delta-function at $\omega$, this is expressed by several point at the discrete points
in the spectrum obtained by the DFT~\cite{harris}.
We study this phenomenon for the AC and WK methods.

\begin{figure}[H]
	\begin{center}
	\includegraphics[width=80mm]{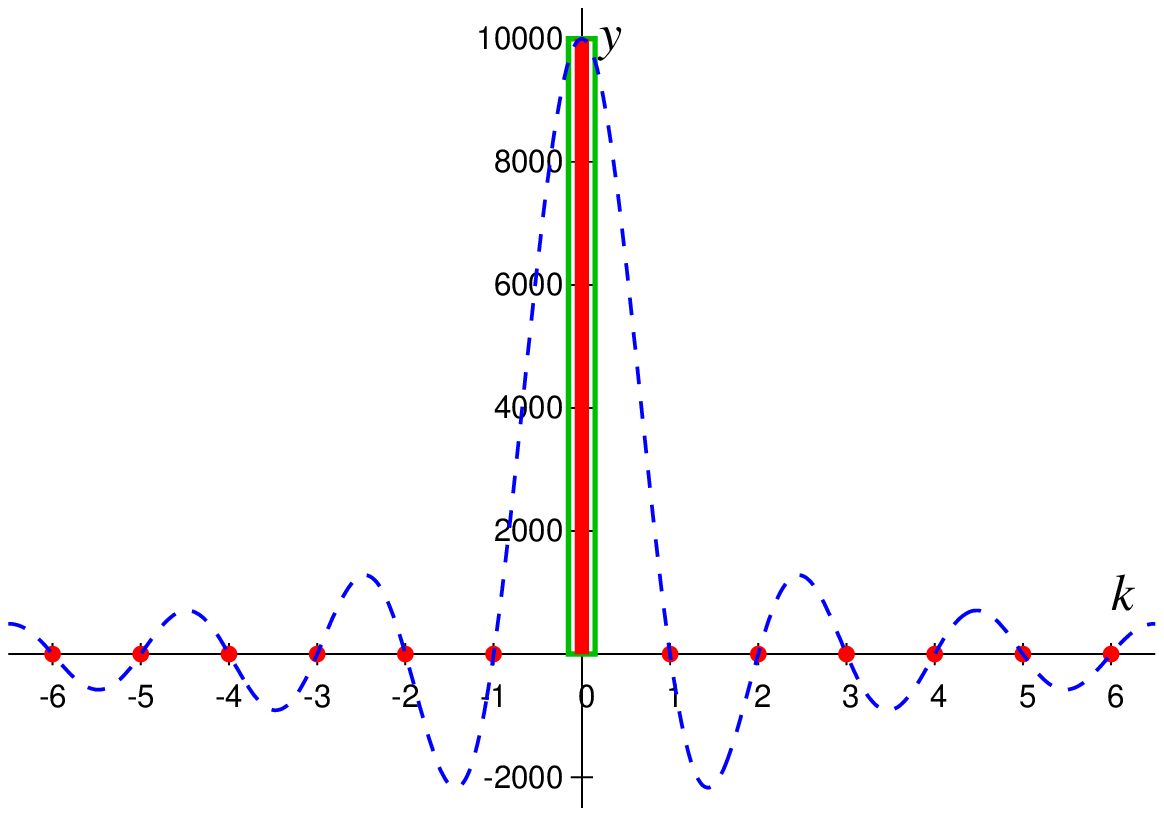}
	\includegraphics[width=80mm]{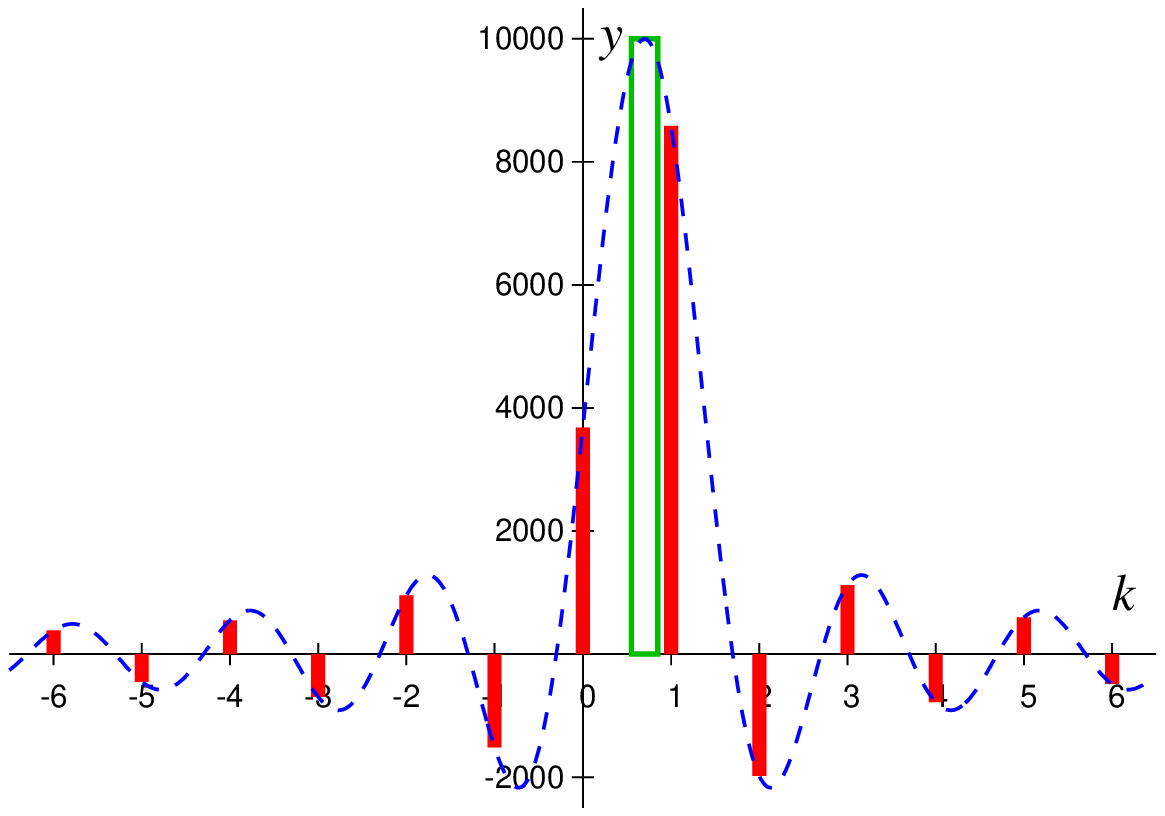}
	\end{center}
\caption{the spectrum of $|\hat{H}(\omega)|$ : $n=10000$, $\Delta=0.5$ : (Left) The true delta-peak is located at $k=0$. (Right) The true delta-peak is located at $k=0.7$.}
	\label{DTFTrec}
\end{figure}

\begin{figure}[H]
	\begin{center}
	\includegraphics[width=80mm]{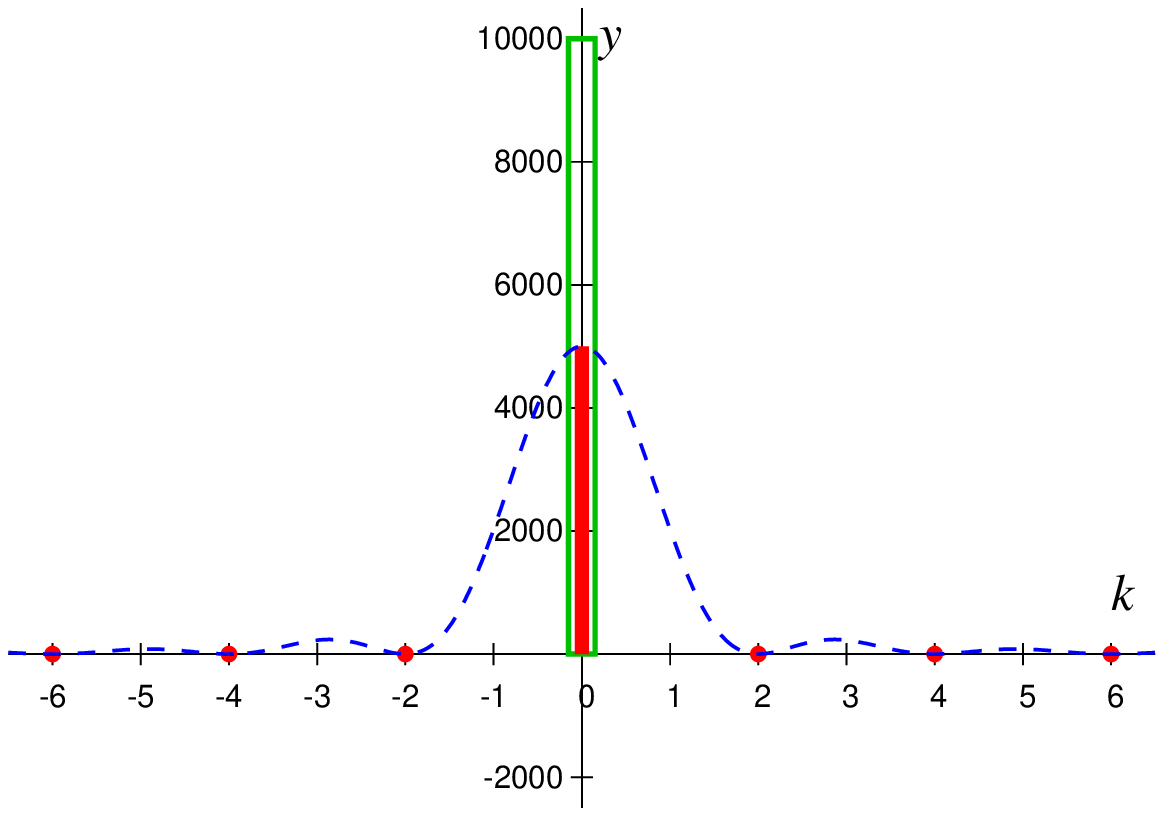}
	\includegraphics[width=80mm]{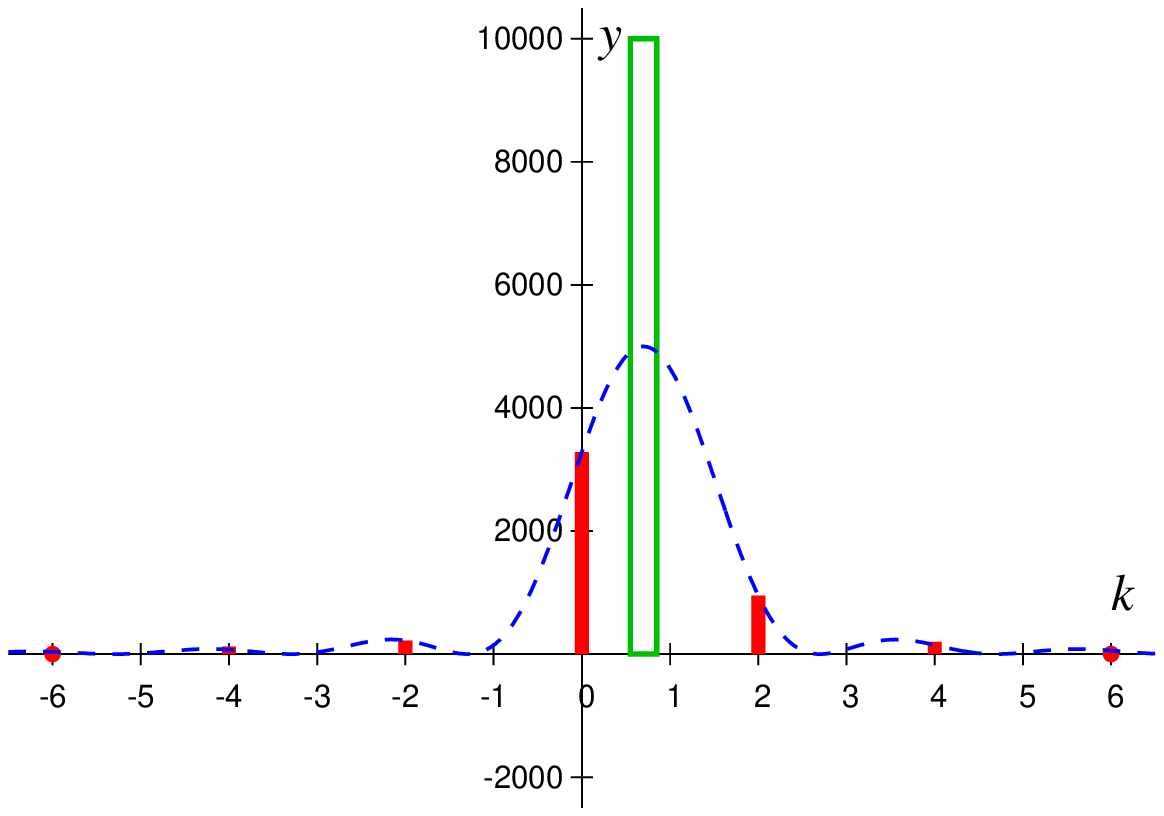}
	\end{center}
\caption{the spectrum of $\frac{1}{4T}|\hat{H}(\omega)|^2$ : $n=10000$, $\Delta=0.5$ : (Left)The true delta-peak is located at $k=0$. (Right) The true delta-peak is located at $k=0.7$.}
	\label{DTFTWK}
\end{figure}

For a time sequence of data $\{f_{k}\}_{k=-\infty}^{\infty}$ such as the autocorrelation function,
the spectrum of the process is given by the DFT:
\begin{eqnarray}
	F(\omega)&=&\Delta\sum_{k=-\infty}^{\infty}f_{k}\mathrm{e}^{-\mathrm{i}\omega k
	\Delta}\left(\simeq\int_{-\infty}^{\infty}f(t)\mathrm{e}^{-\mathrm{i}\omega t}\mathrm{d}t\right),\\
	f_{k}&=&\frac{1}{2\pi}\int_{-\pi/\Delta}^{\pi/\Delta}F(\omega)\mathrm{e}^{\mathrm{i}\omega k
	\Delta}\mathrm{d}\omega,
\end{eqnarray}
where $\Delta$ is the sampling interval.
For a set with a finite number, say $2n$, the transformation reads
\beq
	\widetilde{F}(\omega)\equiv\Delta\sum_{k=-n}^{n-1}f_{k}\mathrm{e}^{-\mathrm{i}\omega
	k\Delta}=\Delta\sum_{k=-\infty}^{\infty}f_{k}h_{k}\mathrm{e}^{-\mathrm{i}\omega k\Delta},
       \label{A0}
\eeq
where
\begin{eqnarray}
	h_{k}=\left\{ \begin{array}{ll}
	1 & \hspace{1cm}k=0,\pm1,\pm2...,\pm(n-1),-n\\
	0 & \hspace{1cm}\mathrm{otherwise.}\\
	 \end{array} \right.
\end{eqnarray}
Expression Eq.~(\ref{A0}) can be written as a convolution:
\beq
	\widetilde{F}(\omega)=\frac{1}{2\pi}\int_{-\pi/\Delta}^{\pi/\Delta}F(\omega')H(\omega-\omega')\mathrm{d}\omega
	=\frac{1}{2\pi}\int_{-\pi/\Delta}^{\pi/\Delta}F(\omega-\omega')H(\omega')\mathrm{d}\omega,
\eeq
where $H(\omega)$ is the DFT of $\{h_{k}\}_{k}$
\begin{eqnarray}
	{H}(\omega)=\Delta\sum_{k=-n}^{n-1}\mathrm{e}^{-\mathrm{i}\omega k\Delta}
	=\Delta \mathrm{e}^{\mathrm{i}\frac{\omega\Delta}{2}}\frac{\mathrm{sin}(\omega n\Delta)}
	{\mathrm{sin}(\omega\Delta/{2})},
\end{eqnarray}
indicating that the spectrum we want is deformed by
\beq
\hat{H}(\omega)\equiv\Delta\frac{\mathrm{sin}(\omega n\Delta)}{\mathrm{sin}(\omega\Delta/2)}.
\label{A1}
\eeq
In Fig.~\ref{DTFTrec}(Left), the blue dashed curve shows the modulus of Eq.~(\ref{A1}).
If $F(\omega)$ is a delta function $\delta(\omega-\omega_{\rm peak})$ the spectrum obtained by DFT is
\beq
\widetilde{F}(\omega)= \frac{1}{2\pi}{H}(\omega-\omega_{\rm peak})
\eeq
for $\omega$ given by the mesh points $\{\omega_k\}$.
If $\omega_{\rm peak}$ is one of the mesh points, the DFT spectrum has a single peak.
However, if $\omega_{\rm peak}$ is located in an interval of mesh points, the DFT spectrum has
several peaks as shown in
Fig.~\ref{DTFTrec}(Right).
The oscillation of $H(\omega)$ even gives peaks with negative amplitude.
The red rods show how the delta function in the correct spectrum at the position of green
box
appears in finite discrete spectrum in the AC and WK methods.
In contrast to the AC method,
in the WK method the spectrum is squared
and the spectrum is given by $|\widetilde{F}(\omega)|^2$, hence the spectrum is positive by construction.
Although Gibbs oscillations are present, they give a width due to the finite time window and do not affect the spectrum much.
In Fig.~\ref{DTFTrec} and \ref{DTFTWK}, we show comparison of $\hat{H}(\omega)$ and $|\hat{H}(\omega)|^2$ respectively.
Note that in the WK method, we can obtain the discrete spectrum only at the points of even $k$'s.

\begin{figure}[H]
	\begin{center}
	\includegraphics[width=80mm]{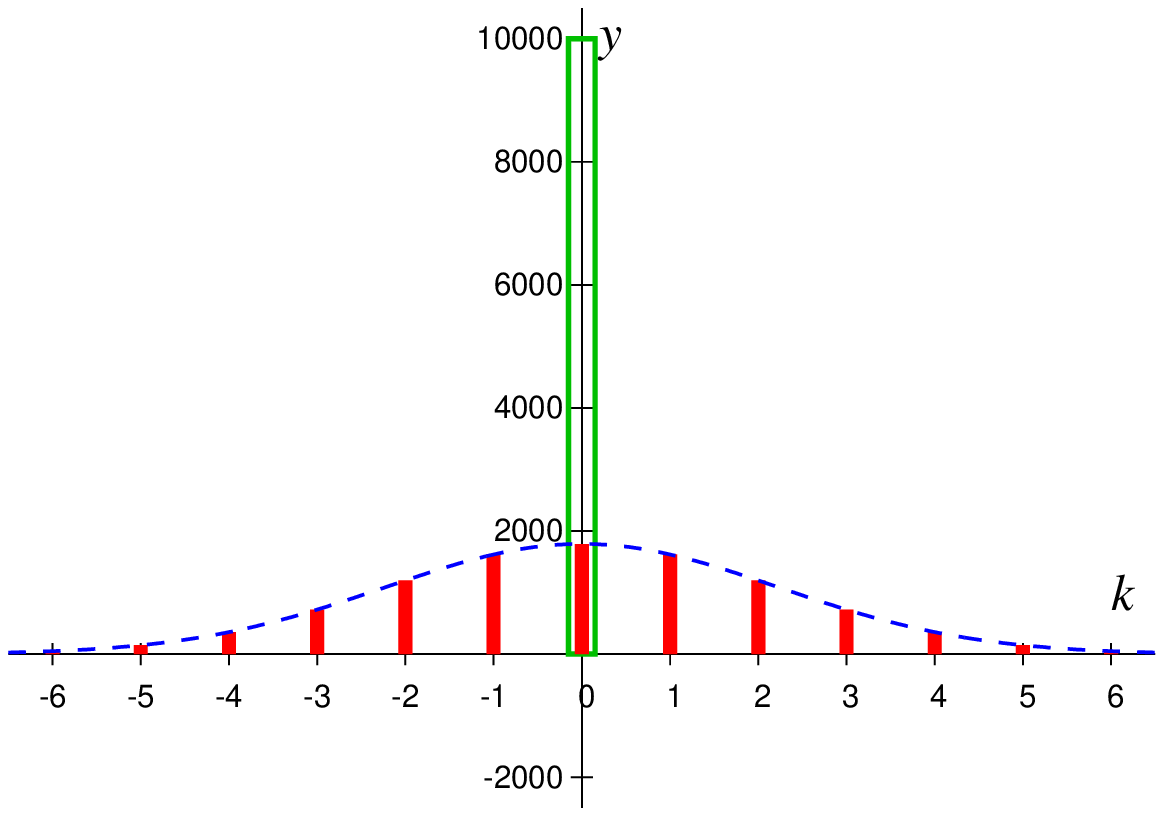}
	\includegraphics[width=80mm]{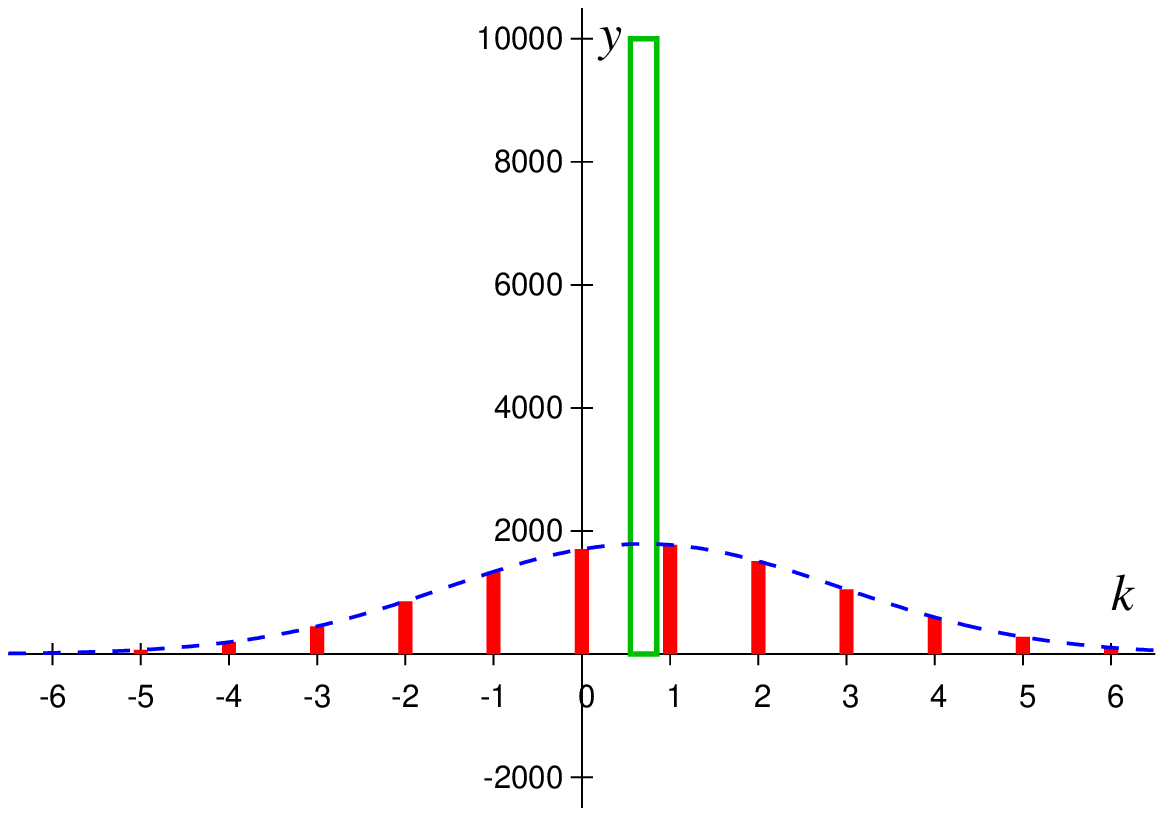}
	\end{center}
\caption{The spectrum of $|G(\omega)|$: $n=10000$, $\Delta=0.5$, and $\alpha=7$.  (Left) The true delta-peak is located at $k=0$. (Right) The true delta-peak is located at $k=0.7$ (green line).}
	\label{DTFTAC}
\end{figure}

In order to avoid these apparent negative peaks, a window function, often a Gaussian, is introduced:
\begin{eqnarray}
	\int_{-T}^{T}f(t)\mathrm{e}^{-\mathrm{i}\omega t}
	\mathrm{e}^{-\frac{1}{2}\left(\alpha\frac{t}{T}\right)^2}\mathrm{d}t
	=2\mathrm{Re}\left[\int_{0}^{T}f(t)\mathrm{e}^{-\mathrm{i}\omega t}
	\mathrm{e}^{-\frac{1}{2}\left(\alpha\frac{t}{T}\right)^2}\mathrm{d}t\right].
\end{eqnarray}
This treatment replaces $f_k$ by
\begin{eqnarray}
	g_{k}=h_{k}\mathrm{e}^{-\frac{1}{2}(\alpha\frac{k}{n})^2},\quad k=0,\pm1,\pm2,...
\end{eqnarray}
The resulting spectrum $|G(\omega)|$ is depicted in Fig.~\ref{DTFTAC}.

The parameter $\alpha$ determines the artificial resolution of the spectrum.
Within this resolution, the Gibbs oscillations are smeared out, hence we can reproduce the spectrum as discussed in section \ref{sec:TDACFN}.

The width of the Gauss window $\alpha$ is known to be taken as
\beq
\mathrm{e}^{-{1\over2}\left(\alpha{n\over n}\right)^2}= \varepsilon \rightarrow \alpha^2=-2\ln \varepsilon,
\eeq
where $\varepsilon$ is a number of the order of the smallest number of the computer resolution,
say  $\varepsilon=10^{-12}$.
\subsection{Conservation of Intensity}
As mentioned in Sec.~\ref{amplitude}, the total amplitude of the spectrum is reproduced correctly by the WK method and the AC method,
despite the presence of Gibbs oscillations caused by performing DFT.
In the subsections that follow, we demonstrate this fact analytically for both methods.
\subsubsection{Wiener-Khinchin method}
In the case where $\omega$ is a continuous variable,
the total amplitude of the spectrum is independent of the time $T$ because of the following relation;
\beq
	\label{cont_WK}
	\frac{1}{T}\int_{-\infty}^{\infty}\left|\frac{1}{2\pi}\int_{0}^{T}\mathrm{e}^{-\mathrm{i}\omega t}\mathrm{d}t\right|^2\mathrm{d}\omega
	=\frac{1}{T}\int_{-\infty}^{\infty}\left(\frac{\mathrm{sin}(\frac{\omega T}{2})}{\pi\omega}\right)^2\mathrm{d}\omega=\frac{1}{2\pi}.
\eeq
In numerical calculations, the Fourier transform is discretized and the Gibbs oscillations
may seem to bring some deviation of the intensity.
But in reality, as shown below, a sum rule such as Eq.~(\ref{cont_WK})
holds even in the discrete case and the correct intensity can be calculated without being affected by the Gibbs oscillations,
independently of the time $T$.

Suppose that there is a peak between two successive $\omega_k$ values.
Then the DFT and the total amplitude of the spectrum are given by
\begin{eqnarray}
	& \displaystyle \frac{1}{T}\sum_{n=-M}^{M}\left|\frac{1}{2\pi}\sum_{k=0}^{N-1}\mathrm{e}^{-\mathrm{i}\omega_n kT/N}\times\frac{T}{N}\right|^2\times\frac{2\pi}{T}
	=\frac{1}{2\pi N^2}\sum_{n=-M}^{M}\left|\frac{\mathrm{sin}\left(\frac{\omega_n T}{2}\right)}{\mathrm{sin}\left(\frac{\omega_n T}{2N}\right)}\right|^2,\\
	& \displaystyle \omega_n=\frac{2\pi n}{T}+\delta,\quad 0\le\delta<\frac{2\pi}{T},
\end{eqnarray}
where $\delta$ is the deviation of the peak's position from a discretized $\omega$, and $M$ should be taken sufficiently large to sum up the amplitude around the peak. At the same time we assume that $M\ll N$. Then we can use the following relations:
\begin{eqnarray}
\label{kinji1}
\mathrm{sin}\left(\frac{\omega_n T}{2N}\right)&=&\mathrm{sin}\left(\frac{\pi n}{N}+\frac{T\delta}{2N}\right)\simeq\frac{\pi n}{N}+\frac{T\delta}{2N},\quad n\in[-M,M].\\
\label{kinji2}
\mathrm{sin}\left(\frac{\omega_n T}{2}\right)&=&\mathrm{sin}\left(\pi n+\frac{T\delta}{2}\right)=(-1)^n\mathrm{sin}\left(\frac{T\delta}{2}\right).
\end{eqnarray}
Thus the sum is given by
\begin{eqnarray}
	\frac{1}{T}\sum_{n=-M}^{M}\left|\frac{1}{2\pi}\sum_{k=0}^{N-1}\mathrm{e}^{-\mathrm{i}\omega_n kT/N}\times\frac{T}{N}\right|^2\times\frac{2\pi}{T}
	&\simeq&\frac{1}{2\pi N^2}\sum_{n=-M}^{M}\left(\frac{(-1)^n\mathrm{sin}\left(\frac{T\delta}{2}\right)}{\frac{\pi n}{N}+\frac{T\delta}{2N}}\right)^2
	\simeq\frac{\mathrm{sin}^2\left(\frac{T\delta}{2}\right)}{2\pi^3}\sum_{n=-\infty}^{\infty}\frac{1}{\left(n+\frac{T\delta}{2\pi}\right)^2}\\
	&=&\frac{1}{2\pi^3}\mathrm{sin}^2\left(\frac{T\delta}{2}\right)\times\left(\pi\mathrm{cosec}\left(\frac{T\delta}{2}\right)\right)^2
	=\frac{1}{2\pi},
\end{eqnarray}
where we use the formula of infinite series $\sum_{n=-\infty}^{\infty}\frac{1}{(n+x)^2}=\left(\pi\mathrm{cosec}(\pi x)\right)^2$\cite{table}.
This result is completely in accord with the continuous one Eq.~(\ref{cont_WK}).

\subsubsection{autocorrelation method}
We can also derive a similar result for the AC method.
In the case where $\omega$ is a continuous variable,
the total amplitude of the spectrum is independent of the time $T$ because of the following relation:
\begin{eqnarray}
	\label{cont_AC}
	\int_{-\infty}^{\infty}2\mathrm{Re}\left[\int_{0}^{T}\mathrm{e}^{-\mathrm{i}\omega t}\mathrm{d}t\right]\mathrm{d}\omega
	=2\pi\int_{-\infty}^{\infty}\frac{\mathrm{sin}(\omega T)}{\pi\omega}\mathrm{d}\omega
	=2\pi.
\end{eqnarray}
This sum rule holds even in the discrete case as shown below.
The basic idea of the derivation is almost the same as in the WK method,
but it should be noted that $\mathrm{d}\omega$ is half of that in the WK method because the time $T$ is essentially two times larger in this case.
The discrete version of Eq.~(\ref{cont_AC}) is given by
\begin{eqnarray}
	\sum_{n=-M}^{M}2\mathrm{Re}\left[\sum_{k=0}^{N-1}\mathrm{e}^{-\mathrm{i}\omega_n kT/N}\times\frac{T}{N}\right]\times\frac{\pi}{T}
	&=&\sum_{n=-M}^{M}2\mathrm{Re}\left[\mathrm{exp}\left(-\mathrm{i}\omega_n\frac{(N-1)T}{2N}\right)\right]\frac{\mathrm{sin}\left(\frac{\omega_nT}{2}\right)}{\mathrm{sin}\left(\frac{\omega_nT}{2N}\right)}\times\frac{\pi}{N}\\
	&=&\frac{2\pi}{N}\sum_{n=-M}^{M}\mathrm{cos}\left(\omega_n\frac{(N-1)T}{2N}\right)\frac{\mathrm{sin}\left(\frac{\omega_nT}{2}\right)}{\mathrm{sin}\left(\frac{\omega_nT}{2N}\right)},
\end{eqnarray}
where
\begin{eqnarray}
	\omega_n&=&\frac{\pi n}{T}+\delta,\quad 0\le\delta<\frac{\pi}{T}.
\end{eqnarray}
Using the following relations:
\begin{eqnarray}
\mathrm{cos}\left(\omega_n\frac{(N-1)T}{2N}\right)
	&=&\mathrm{cos}\left(\frac{\omega_n T}{2}\right)\mathrm{cos}\left(\frac{\omega_n T}{2N}\right)+\mathrm{sin}\left(\frac{\omega_n T}{2}\right)\mathrm{sin}\left(\frac{\omega_n T}{2N}\right),\\
\label{tan}
\mathrm{tan}\left(\frac{\omega_n T}{2N}\right)&\simeq&\frac{\omega_n T}{2N},\quad n\in[-M,M],
\quad {\rm for}\quad M\ll N,
\end{eqnarray}
and
\begin{eqnarray}
\mathrm{sin}\left(\frac{\omega_n T}{2}\right)&=&\mathrm{sin}\left(\pi n+T\delta\right)
	=\mathrm{sin}(\pi n)\mathrm{cos}(T\delta)+\mathrm{cos}(\pi n)\mathrm{sin}(T\delta)
	=(-1)^n\mathrm{sin}(T\delta),
\end{eqnarray}
the sum is given by
\begin{eqnarray}
	\sum_{n=-M}^{M}2\mathrm{Re}\left[\sum_{k=0}^{N-1}\mathrm{e}^{-\mathrm{i}\omega_n kT/N}\times\frac{T}{N}\right]\times\frac{2\pi}{T}
	&=&\frac{2\pi}{N}\sum_{n=-M}^{M}\left\{\mathrm{cos}\left(\frac{\omega_n T}{2}\right)\mathrm{cos}\left(\frac{\omega_n T}{2N}\right)+\mathrm{sin}\left(\frac{\omega_n T}{2}\right)\mathrm{sin}\left(\frac{\omega_n T}{2N}\right)\right\}\frac{\mathrm{sin}\left(\frac{\omega_nT}{2}\right)}{\mathrm{sin}\left(\frac{\omega_nT}{2N}\right)}\\
	&=&\frac{2\pi}{N}\sum_{n=-M}^{M}\frac{\mathrm{sin}\left(\frac{\omega_n T}{2}\right)\mathrm{cos}\left(\frac{\omega_n T}{2}\right)}{\mathrm{tan}\left(\frac{\omega_n T}{2N}\right)}+\mathcal{O}\left(\frac{M}{N}\right)
	\simeq2\pi\sum_{n=-M}^{M}\frac{\mathrm{sin}\left(\frac{\omega_n T}{2}\right)\mathrm{cos}\left(\frac{\omega_n T}{2}\right)}{\frac{\omega_n T}{2}}\\
	&\simeq&2\pi\sum_{n=-\infty}^{\infty}\frac{\mathrm{sin}\left(\frac{\omega_n T}{2}\right)\mathrm{cos}\left(\frac{\omega_n T}{2}\right)}{\frac{\omega_n T}{2}}
	=2\pi\sum_{n=-\infty}^{\infty}\frac{\mathrm{sin}\left(\omega_n T\right)}{\omega_n T}\\
	&=&2\mathrm{sin}\left(\delta T\right)\sum_{n=-\infty}^{\infty}\frac{(-1)^n}{n+\frac{T\delta}{\pi}}
	=2\mathrm{sin}\left(\delta T\right)\times\pi\mathrm{cosec}\left(T\delta\right)=2\pi,
\end{eqnarray}
where we use the formula of infinite series $\sum_{n=-\infty}^{\infty}\frac{(-1)^n}{n+x}=\pi\mathrm{cosec}(\pi x)$\cite{table}.
%
%
\section{Statistical properties of the thermal typical state approach}\label{sec1}
The underlying idea of the thermal typical state approach is that it is a good approximation to replace
$\mathbf{Tr\;} X$ by $\langle\Phi|X|\Phi\rangle$ where $X=X^\dagger$ is a $D\times D$ Hermitian matrix,
the random state $|\Phi\rangle=\sum_{a=}^D \xi_a |a\rangle$ where
the $\xi_a$'s are complex-valued Gaussian random variables and the set $\{|a\rangle\}$ can be any complete
set of orthonormal states for the Hilbert space of dimension $D$.
The demonstration that this approximation is indeed useful requires a proof that
by averaging over the $\xi_a$ we recover the correct answer $\mathbf{Tr\;} X$
and that the variance of $\langle\Phi|X|\Phi\rangle$ is bounded.
\subsection{Thermal typical state}
Let us put $X=\mathrm{e}^{-\beta\mathcal{H}/2} Y \mathrm{e}^{-\beta\mathcal{H}/2}$ where $H=H^\dagger$ and $Y=Y^\dagger$
which implies $X=X^\dagger$.
Using Eqs.~(\ref{RS31text})--~(\ref{RS34text}) we have
\begin{eqnarray}
\mathrm{E}[\langle\Phi|X|\Phi\rangle]
&=&
2\sigma^2 \mathbf{Tr\;} X = 2\sigma^2 \mathbf{Tr\;} \mathrm{e}^{-\beta\mathcal{H}} Y
,
\label{RS36}
\end{eqnarray}
and
\begin{eqnarray}
\mathrm{Var}\left(\langle\Phi|X|\Phi\rangle\right)
&=&4\sigma^4 \mathbf{Tr\;} \mathrm{e}^{-\beta\mathcal{H}/2} Y \mathrm{e}^{-\beta\mathcal{H}/2} \mathrm{e}^{-\beta\mathcal{H}/2} Y \mathrm{e}^{-\beta\mathcal{H}/2}
\nonumber \\
&=&4\sigma^4 \mathbf{Tr\;} \left(\mathrm{e}^{-\beta\mathcal{H}} Y\right)\left(\mathrm{e}^{-\beta\mathcal{H}} Y\right)
=4\sigma^4 \mathbf{Tr\;} \left(\mathrm{e}^{-\beta\mathcal{H}} Y\right)^2
.
\label{RS37}
\end{eqnarray}
In our simulations, we use the state defined by
\begin{eqnarray}
|\Phi_\beta\rangle &\equiv& \mathrm{e}^{-\beta\mathcal{H}/2}|\Phi\rangle
,
\label{RS40}
\end{eqnarray}
to compute estimates of thermal equilibrium expectation values. So we can write $\langle\Phi_{\beta}|Y|\Phi_{\beta}\rangle=\langle\Phi|X|\Phi\rangle$.
With $A=\mathrm{e}^{-\beta\mathcal{H}}$ and $B=Y$ we can derive the upper bound of the variance as following;
\begin{eqnarray}
\mathrm{Var}^2\left(\langle\Phi_{\beta}|Y|\Phi_{\beta}\rangle\right)
&=&16\sigma^8 \left|\mathbf{Tr\;} (AB)^2\right|^2=\left|((AB)^\dagger,AB)\right|^2
\nonumber \\
&\le& (AB,AB) ((AB)^\dagger,(AB)^\dagger) = \left(\mathbf{Tr\;} (AB)^\dagger AB \right)^2
= \left(\mathbf{Tr\;} A^2B^2\right)^2
,
\label{RS38}
\end{eqnarray}
where we used the Schwarz inequality because $\mathbf{Tr\,} C^\dagger D\equiv (C,D) $ defines a scalar product in general.

As $\mathbf{Tr\;} (AB)^2=\mathbf{Tr\;} (\mathrm{e}^{-\beta\mathcal{H}/2} Y \mathrm{e}^{-\beta\mathcal{H}/2})(\mathrm{e}^{-\beta\mathcal{H}/2} Y \mathrm{e}^{-\beta\mathcal{H}/2})^\dagger\ge0$, it follows that
\begin{center}
\parbox[t]{0.9\hsize}{%
\begin{eqnarray}
\mathrm{Var}\left(\langle\Phi_{\beta}|Y|\Phi_{\beta}\rangle\right)
&=&4\sigma^4 \mathbf{Tr\;} \left(\mathrm{e}^{-\beta\mathcal{H}} Y\right)^2
\le 4\sigma^4
\mathbf{Tr\;} \mathrm{e}^{-2\beta\mathcal{H}}Y^2
\label{RS39a}
\\
\mathrm{RSD}^2(\langle\Phi_{\beta}|Y|\Phi_{\beta}\rangle)&\le&
\frac{\mathbf{Tr\;} \mathrm{e}^{-2\beta\mathcal{H}} Y^2}{\left(\mathbf{Tr\;} \mathrm{e}^{-\beta\mathcal{H}} Y\right)^2}
.
\label{RS39}
\end{eqnarray}
}
\end{center}
\subsection{Bounds on the variance of the estimate of the partition function}\label{RS}

The inequality Eq.~(\ref{RS39}) is very useful
to prove that the statistical error on the estimate of the partition function
vanishes exponentially with the system size.
Specializing to $Y=\openone$ and noting that
$\langle\Phi_{\beta}|\Phi_{\beta}\rangle>0$ we have
\begin{eqnarray}
\mathrm{E}[\langle\Phi_{\beta}|\Phi_{\beta}\rangle]
&=&2\sigma^2 \mathbf{Tr\;} \mathrm{e}^{-\beta\mathcal{H}} =2\sigma^2 Z(\beta)=2\sigma^2 \mathrm{e}^{-\beta F(\beta)},
\label{RS41a}
\\
\mathrm{Var}\left(\langle\Phi_{\beta}|\Phi_{\beta}\rangle\right)
&=&4\sigma^4 \mathbf{Tr\;} \mathrm{e}^{-2\beta\mathcal{H}}= 4\sigma^4 Z(2\beta)= 4\sigma^4 \mathrm{e}^{-2\beta F(2\beta)},
\label{RS41b}
\\
\mathrm{RSD}(\langle\Phi_{\beta}|\Phi_{\beta}\rangle)
&\le& \mathrm{e}^{-\beta (F(2\beta)-F(\beta))}
,
\label{RS41c}
\end{eqnarray}
where $F(\beta)$ denotes the free energy of the system at the inverse temperature $\beta$.
As the free energy is an extensive quantity, i.e. it is proportional to the number of particles, and
$F(2\beta)- F(\beta)>0$ for $\beta>0$, we have $\mathrm{e}^{-\beta (F(2\beta)-F(\beta))}={\cal O}(\mathrm{e}^{-N})$
and we obtain
\begin{eqnarray}
\mathrm{RSD}(\langle\Phi_{\beta}|\Phi_{\beta}\rangle)
&\le&\mathrm{e}^{-\beta (F(2\beta)-F(\beta))}={\cal O}(\mathrm{e}^{-N})
.
\label{RS42}
\end{eqnarray}
On the other hand, the justification of above estimation may not be obvious for the infinite temperature $\beta\rightarrow0$.
But still the convergence of the partition function is guaranteed even in this case because by using
\begin{eqnarray}
\lim_{\beta\rightarrow0} \mathrm{e}^{-\beta F(\beta)} &=& \mathbf{Tr\,} \openone = D
,
\label{RS43}
\end{eqnarray}
we find that
\begin{eqnarray}
\lim_{\beta\rightarrow0}\mathrm{RSD}
\left(\langle\Phi_{\beta}|\Phi_{\beta}\rangle\right)
&\le&\frac{1}{\sqrt{D}},
\label{RS44}
\end{eqnarray}
in agreement with \cite{hams}.
Moreover,
by the Schwarz inequality we have in general
\begin{eqnarray}
|\mathbf{Tr\,} X|^2=|(\openone,X)|^2&\le& (\openone,\openone)(X,X)=D \mathbf{Tr\,} X^\dagger X
,
\label{RS35atext}
\end{eqnarray}
and hence
\begin{eqnarray}
\mathrm{RSD}(\langle\Phi_{\beta}|\Phi_{\beta}\rangle)
=\mathrm{RSD}(\langle\Phi|X|\Phi\rangle) \ge \frac{1}{\sqrt{D}}
\label{RS35text}
\end{eqnarray}
from the definition of $\mathrm{RSD}$ given by Eq.~(\ref{RS34text}).
Thus, it follows that
\begin{eqnarray}
\lim_{\beta\rightarrow0}\mathrm{RSD}
\left(\langle\Phi_{\beta}|\Phi_{\beta}\rangle\right)
&=&\frac{1}{\sqrt{D}}.
\label{RS44a}
\end{eqnarray}
From Eqs.~(\ref{RS42}) and~(\ref{RS44a})
we can conclude that $\mathrm{RSD}\left(\langle\Phi_{\beta}|\Phi_{\beta}\rangle\right)$
vanishes exponentially with the system size $N$.
Recall that we can choose $\sigma$ as we like.
From Eq.~(\ref{RS41a}), it is clear that if we choose $\sigma=1/\sqrt{2}$ we have
$\mathrm{E}[\langle\Phi_{\beta}|\Phi_{\beta}\rangle]=\mathbf{Tr\;} e^{-\beta\mathcal{H}} = \mathrm{e}^{-\beta F(\beta)}$
that is, the norm of the thermal state is, up to statistical fluctuations which vanish as $1/D$,
equal to the partition function.

\subsection{Approximate estimates}\label{AE}
In this subsection we write $Z=\mathrm{e}^{-\beta\mathcal{H}}$ and to simplify the writing (but without loss of generality), we also choose $\sigma=1/\sqrt{2}$.
The general idea of the approach is that it suffices to generate one thermal typical state
$|\Phi_{\beta}\rangle$ to find good estimates for
$\langle Y\rangle_{\mathrm{eq}}=\mathbf{Tr\;} \mathrm{e}^{-\beta\mathcal{H}} Y/\mathbf{Tr\;} \mathrm{e}^{-\beta\mathcal{H}}$
The question is if we can prove that the statistical fluctuations are small in some sense.
The problem is the following:
In the simulation we generate a thermal typical state $|\Phi_{\beta}\rangle$
and compute $\langle\Phi_{\beta}|Y|\Phi_{\beta}\rangle/\langle\Phi_{\beta}|\Phi_{\beta}\rangle$.
In the present paper, we assume that the denominator and numerator are averaged and then we form the ratio.
But if we take the ratio first and then we take the average, the estimate is different.
In the following we study the variance in this case.
Although from Eq.~(\ref{RS39a}) we know that variances of these two quantities
are bounded from above, we do not yet have a bound on the variance of the ratio of them.
The purpose of this subsection is to address this point.

A simple method to estimate averages and variances of the ratio is to make
use of the multivariate Taylor expansion for the average
\begin{eqnarray}
\mathrm{E}\left[
\frac{x}{y}\right]
\approx
\frac{\mathrm{E}[x]}{\mathrm{E}[y]}-\frac{\mathrm{Cov}[x,y]}{\mathrm{E}^2[y]}+\frac{\mathrm{E}[x]\mathrm{Var}[y]}{\mathrm{E}^3[y]}
,
\label{AE0}
\end{eqnarray}
where $\mathrm{Cov}[x,y]=\mathrm{E}[xy]-\mathrm{E}[x]\mathrm{E}[y]$
and a similar approximation for the variance
\begin{eqnarray}
\mathrm{Var}\left[\frac{x}{y}\right]
\approx
\frac{\mathrm{Var}[x]}{\mathrm{E}^2[y]}-2\frac{\mathrm{E}[x]\mathrm{Cov}[x,y]}{\mathrm{E}^3[y]}+\frac{\mathrm{E^2}[x]\mathrm{Var}[y]}{\mathrm{E}^4[y]}
.
\label{AE1}
\end{eqnarray}
Now we generate a thermal typical state $|\Phi_{\beta}\rangle$ and compute ratios
\begin{eqnarray}
R(Y,Z)&\equiv&\frac{\langle\Phi_{\beta}|Y|\Phi_{\beta}\rangle}{\langle\Phi_{\beta}|\Phi_{\beta}\rangle}
.
\label{AE2}
\end{eqnarray}
To calculate its average and variance according to Eqs.~(\ref{AE0}) and (\ref{AE1}), let us prepare the covariance $\mathrm{Cov}[\langle\Phi_{\beta}|Y|\Phi_{\beta}\rangle,\langle\Phi_{\beta}|\Phi_{\beta}\rangle]$:
\begin{eqnarray}
\mathrm{E}[\langle\Phi_{\beta}|Y|\Phi_{\beta}\rangle\langle\Phi_{\beta}|\Phi_{\beta}\rangle]
&=&\sum_{a,p,b,q=1}^D \mathrm{E}[ \xi_a^\ast \xi_p^{\phantom{\ast}} \xi_b^\ast \xi_q^{\phantom{\ast}} ]
\langle a|ZY|p\rangle \langle b|Z|q\rangle
=\mathbf{Tr\;} Z^2 Y + \left(\mathbf{Tr\;}ZY \right)\left(\mathbf{Tr\;} Z\right),\\
\mathrm{Cov}[\langle\Phi_{\beta}|Y|\Phi_{\beta}\rangle,\langle\Phi_{\beta}|\Phi_{\beta}\rangle]
&=&\mathrm{E}[\langle\Phi_{\beta}|Y|\Phi_{\beta}\rangle\langle\Phi_{\beta}|\Phi_{\beta}\rangle] -
\mathrm{E}[\langle\Phi_{\beta}|Y|\Phi_{\beta}\rangle]\mathrm{E}[\langle\Phi_{\beta}|\Phi_{\beta}\rangle]
=\mathbf{Tr\;} Z^2 Y.
\label{AV3}
\end{eqnarray}
By using this relation, it follows that
\begin{eqnarray}
\mathrm{E}\left[
\frac{\langle\Phi_{\beta}|Y|\Phi_{\beta}\rangle}{\langle\Phi_{\beta}|\Phi_{\beta}\rangle}
\right]
&\approx&                                 \frac{\mathbf{Tr\;} Z Y}{\mathbf{Tr\;} Z}
-
\frac{\mathbf{Tr\;} Z^2 Y}{\left(\mathbf{Tr\;} Z\right)^2}
+
\frac{\mathbf{Tr\;} Z Y}{\mathbf{Tr\;} Z}
\frac{\mathrm{Var}[\langle\Phi_{\beta}|\Phi_{\beta}\rangle]}{\left(\mathbf{Tr\;} Z\right)^2}
\nonumber \\
&&=
\frac{\mathbf{Tr\;} Z Y}{\mathbf{Tr\;} Z}
-
\frac{\mathbf{Tr\;} Z^2 Y}{\left(\mathbf{Tr\;} Z\right)^2}
+
\frac{\mathbf{Tr\;} Z Y}{\mathbf{Tr\;} Z}
\frac{\mathbf{Tr\;} Z^2}{\left(\mathbf{Tr\;} Z\right)^2}
.
\label{AE3}
\end{eqnarray}
The second and third term in Eq.~(\ref{AE3}) vanish exponentially
with the system size because we see
\begin{eqnarray}
\frac{\mathbf{Tr\;} Z^2 Y}{\left(\mathbf{Tr\;} Z\right)^2}
\ \&\
\frac{\mathbf{Tr\;} Z Y}{\mathbf{Tr\;} Z}
\frac{\mathbf{Tr\;} Z^2}{\left(\mathbf{Tr\;} Z\right)^2}
\ \le\
\Vert Y\Vert
\frac{\mathbf{Tr\;} Z^2}{\left(\mathbf{Tr\;} Z\right)^2}
=\Vert Y\Vert\mathrm{e}^{-2\beta (F(2\beta)-F(\beta))}
\end{eqnarray}
where $\Vert Y\Vert=\max_\psi |\langle\psi| Y |\psi\rangle|$ is the largest (in absolute value) eigenvalue of $Y$.
Hence for large $D$ we have
\begin{eqnarray}
\mathrm{E}\left[
\frac{\langle\Phi_{\beta}|Y|\Phi_{\beta}\rangle}{\langle\Phi_{\beta}|\Phi_{\beta}\rangle}
\right]
&\approx&
\frac{\mathbf{Tr\;} Z Y}{\mathbf{Tr\;} Z} = \langle Y \rangle_{\mathrm{eq}}
,
\label{AE4}
\end{eqnarray}
as expected.
Similarly, for the variance we have
\begin{eqnarray}
\mathrm{Var}\left[
\frac{\langle\Phi_{\beta}|Y|\Phi_{\beta}\rangle}{\langle\Phi_{\beta}|\Phi_{\beta}\rangle}
\right]
&\approx&
\frac{\mathrm{Var}[\langle\Phi_{\beta}|Y|\Phi_{\beta}\rangle]}{\left(\mathbf{Tr\;} Z\right)^2}
-2
\frac{\mathbf{Tr\;} Z Y}{\mathbf{Tr\;} Z}
\frac{\mathbf{Tr\;} Z^2 Y}{\left(\mathbf{Tr\;} Z\right)^2}
+
\left(\frac{\mathbf{Tr\;} Z Y}{\mathbf{Tr\;} Z}\right)^2
\frac{\mathrm{Var}[\langle\Phi_{\beta}|\Phi_{\beta}\rangle]}{\left(\mathbf{Tr\;} Z\right)^2}
\nonumber \\
&&=
\frac{\mathbf{Tr\;} (Z Y)^2}{\left(\mathbf{Tr\;} Z\right)^2}
-2
\frac{\mathbf{Tr\;} Z Y}{\mathbf{Tr\;} Z}
\frac{\mathbf{Tr\;} Z^2 Y}{\left(\mathbf{Tr\;} Z\right)^2}
+
\left(\frac{\mathbf{Tr\;} Z Y}{\mathbf{Tr\;} Z}\right)^2
\frac{\mathbf{Tr\;} Z^2}{\left(\mathbf{Tr\;} Z\right)^2}
\nonumber \\
&&=
\frac{\mathbf{Tr\;} Z^2}{\left(\mathbf{Tr\;} Z\right)^2}
\bigg\{
\frac{\mathbf{Tr\;} (Z Y)^2}{\mathbf{Tr\;} Z^2}
-2
\frac{\mathbf{Tr\;} Z Y}{\mathbf{Tr\;} Z}
\frac{\mathbf{Tr\;} Z^2 Y}{\mathbf{Tr\;} Z^2}
+
\left(\frac{\mathbf{Tr\;} Z Y}{\mathbf{Tr\;} Z}\right)^2
\bigg\}
,\label{AE5}
\end{eqnarray}
It is easy to find an upper bound for the terms in the curly brackets.
We have
\begin{eqnarray}
\left|
\frac{\mathbf{Tr\;} (Z Y)^2}{\mathbf{Tr\;} Z^2}
+
\left(
\frac{\mathbf{Tr\;} Z Y}{\mathbf{Tr\;} Z}
\right)^2
-2
\frac{\mathbf{Tr\;} Z Y}{\mathbf{Tr\;} Z}
\frac{\mathbf{Tr\;} Z^2 Y}{\mathbf{Tr\;} Z^2}
\right|
&\le&4 \Vert Y\Vert^2
,
\label{AV5}
\end{eqnarray}
and hence we find
\begin{center}
\parbox[t]{0.9\hsize}{%
\begin{eqnarray}
\mathrm{Var}\left[
\frac{\langle\Phi_{\beta}|Y|\Phi_{\beta}\rangle}{\langle\Phi_{\beta}|\Phi_{\beta}\rangle}
\right]&\le&
4\Vert Y\Vert^2
\mathrm{e}^{-2\beta (F(2\beta)-F(\beta))}
,
\label{AV6}
\end{eqnarray}
}
\end{center}
showing that the variance vanishes exponentially with the system size $N$.

%
%


\begin{thebibliography}{99}
\bibitem{discuss} The relation to real measurements is a very interesting problem which will be discussed separately.
\end{thebibliography}
\end{document}